\documentclass[twocolumn,prd,nofootinbib,superscriptaddress,longbibliography]{revtex4} %
\newcommand\ForInternalReference[1]{}
\newcommand\SkipForEarlyCirculation[1]{}

\newcommand\SkipPP[1]{}
\usepackage[colorlinks=true,citecolor=blue,urlcolor=blue]{hyperref}
\usepackage{verbatim}
\usepackage{graphicx}
\usepackage{dcolumn}
\usepackage{bm}
\usepackage{color}
\usepackage{xcolor}
\usepackage{url}
\usepackage{float}
\usepackage{multirow}
\usepackage{adjustbox}
\usepackage[normalem]{ulem}
\usepackage{bm}
\usepackage{placeins}
\usepackage{afterpage}
\usepackage{amsmath,amssymb}
\usepackage{acronym}
\usepackage{booktabs}      % for \toprule, \midrule, \bottomrule
\usepackage{tabularx}      % for tables that automatically fit \linewidth
\usepackage{array}         % (optional) better column formatting

\newcommand\optional[1]{}
\acrodef{NR}[NR]{Numerical Relativity}

\usepackage{color}
\definecolor{amber}{rgb}{1.0, 0.75, 0.0}
\definecolor{orange}{rgb}{1.0, 0.5, 0.0}
\definecolor{amaranth}{rgb}{0.9, 0.17, 0.31}

\graphicspath{{./Figures/}}

\def\ltsima{$\; \buildrel < \over \sim \;$}
\def\simlt{\lower.5ex\hbox{\ltsima}}
\def\gtsima{$\; \buildrel > \over \sim \;$}
\def\simgt{\lower.5ex\hbox{\gtsima}}

\def\eos#1{equation of state#1 (EOS#1)\gdef\eos{EOS}}
\def\QNM#1{quasi-normal mode#1 (QNM#1)\gdef\QNM{QNM}}
\def\ns#1{neutron star#1 (NS#1)\gdef\ns{NS}}
\def\gw#1{gravitational wave#1 (GW#1)\gdef\gw{GW}}
\def\bh#1{black hole#1 (BH#1)\gdef\bh{BH}}
\def\bbh#1{binary black hole#1  (BBH#1)\gdef\bbh{BBH}}
\def\bns#1{binary neutron star#1 (BNS#1)\gdef\bns{BHS}}
\def\bhns#1{black hole - neutron star#1 (BHNS#1)\gdef\bhns{BHNS}}
\def\nsbh#1{neutron star - black hole#1 (NSBH#1)\gdef\nsbh{NSBH}}
\def\nr#1{numerical relativity#1 (NR#1)\gdef\nr{NR}}

\definecolor{burntorange}{HTML}{CC6600}

\newcommand{\UT}{\affiliation{Center for Gravitational Physics, The University of Texas at Austin, Austin, Texas 78712, USA}}

\begin{document}

\title{Toward a complete understanding of the properties of GW190521} 
\author{Aasim Jan}
\UT
\affiliation{TAPIR, California Institute of Technology, Pasadena, CA 91125, USA}
\affiliation{LIGO Laboratory, California Institute of Technology, Pasadena, CA 91125, USA}
\author{Sophia Nicolella}
\UT
\author{Aaron Zimmerman}
\UT
\author{Deirdre Shoemaker}
\UT
\author{Katerina Chatziioannou}
\affiliation{TAPIR, California Institute of Technology, Pasadena, CA 91125, USA}
\affiliation{LIGO Laboratory, California Institute of Technology, Pasadena, CA 91125, USA}
\author{Richard O'Shaughnessy}
\affiliation{Center for Computational Relativity and Gravitation, Rochester Institute of Technology, Rochester, New York 14623, USA}
\author{Muhammed Saleem}
\UT

%+++++++++++++++++++++++++++++++++++++++++++++++++++++++++++++++++++++++++++++++++++++++++++++++++++
\begin{abstract}

Interpreting GW190521 remains challenging due to its exceptionally high total mass and correspondingly small number of observable gravitational-wave cycles, which gives rise to degeneracies between different physical interpretations. Motivated by the outstanding questions surrounding the event, we present a comprehensive reanalysis of GW190521 using waveform models drawn from multiple families, including a model that includes both eccentricity and spin-precession. 
We find evidence that the source was highly eccentric, with an inferred eccentricity of $0.81^{+0.05}_{-0.01}$ at $5\,$Hz, and show that non-eccentric interpretations arise from failing to probe the high-likelihood region at large eccentricities. 
We further demonstrate that when eccentricity is included in the analysis, there is no indication of spin precession, suggesting that the apparent precession in quasicircular analyses is at least partially driven by an eccentricity--spin-precession degeneracy.
While all waveform models favor a comparable-mass binary, one model admits a secondary mode at more asymmetric masses. We find a robust probability $(\geq97\%)$ that the primary black hole lies within the pair-instability supernova mass gap ($\sim60\text{--}130\,M_\odot$), whereas the probability that the secondary lies in the gap depends on the waveform model. Through inference of GW190521-like injections, we show that the waveform systematics observed in the analysis of GW190521 could arise from the response of models to unmodeled eccentricity, rather than from intrinsic differences between waveform models. When analyzing a quasicircular spin-precessing injection similar to GW190521, we find no significant model dependence in the inferred parameters. Finally, because the inferred eccentricity lies beyond the calibration regime of current models, we test its robustness by recovering highly eccentric numerical-relativity waveforms.

\end{abstract}
\maketitle
%+++++++++++++++++++++++++++++++++++++++++++++++++++++++++++++++++++++++++++++++++++++++++++++++++++
\section{Introduction}

On May 21, 2019, the Advanced LIGO \cite{Aasi_2015} and Advanced Virgo \cite{Acernese_2015} observatories detected the gravitational-wave event GW190521 \cite{PhysRevLett.125.101102,Abbott_2020}, which remained the most massive binary black hole merger reported by the LIGO--Virgo--KAGRA (LVK) Collaboration until the detection of GW231123 \cite{theligoscientificcollaboration2025gwtc40updatinggravitationalwavetransient,Abac_2025}, excluding candidates identified by independent search pipelines \cite{Zackay_2019,Venumadhav_2020,Olsen_2022,Mehta_2025,Nitz_2019_ogc,Nitz_2020_ogc,Nitz_2021_ogc,Nitz_2023_ogc,wadekar2025newblackholemergers,Koloniari_2025}.
The analysis by the collaboration inferred exceptionally large source-frame component masses of $85^{+21}_{-14}\,M_\odot$ and $66^{+17}_{-18}\,M_\odot$, with the primary black hole lying within the theorized pair-instability supernova (PISN) mass gap ($\sim60\text{--}130\,M_\odot$) \cite{Farmer_2019,Farmer_2020,Woosley_2021,hendriks2023pulsationalpairinstabilitysupernovaegravitationalwave} with high credibility. 
The merger remnant was estimated to have a mass of $142^{+28}_{-16}\,M_\odot$, providing the first strong observational evidence for the existence of an intermediate-mass black hole, a class of black holes with masses in the range $10^2\text{--}10^3\,M_\odot$ \cite{Ebisuzaki_2001,Greene_2020_imbh,mezcua_imbh,koliopanos2018intermediatemassblackholes}. 
Furthermore, the signal showed mild evidence for spin-induced precession. 
Given the inferred properties of the system, the LVK analysis favored a dynamical formation scenario over isolated binary evolution, with the inferred properties being consistent with a stellar-merger scenario in dense star clusters \cite{Abbott_2020}. 
These characteristics make GW190521 an important probe of black-hole formation physics.

However, due to the exceptionally high mass of the system, the observed signal contains only a few cycles within the sensitive frequency band of the detectors, making the inference particularly sensitive to waveform-modeling assumptions and to degeneracies between eccentricity and spin precession \cite{Ramos_Buades_2020,Romero_Shaw_2023,Xu_2023,Divyajyoti_2024,divyajyoti2025biasedparameterinferenceeccentric,PhysRevLett.126.201101}. 
Furthermore, given the possibility that the system formed through a dynamical formation channel, the signal could exhibit signatures of orbital eccentricity \cite{Rodriguez_2018,Gondan_2019,PhysRevD.97.103014,Zevin_2019}. 
The initial LVK analysis employed quasicircular waveform models, leaving open the possibility that eccentricity contributes to the observed signal morphology.

Numerous studies have investigated whether GW190521 exhibits signatures of orbital eccentricity, reaching substantially different conclusions. 
The analysis in Ref.~\cite{Romero-Shaw_2020}, using a quadrupole-only waveform model restricted to eccentricities below $0.2$, found strong support for an eccentric aligned-spin interpretation over a quasicircular aligned-spin interpretation and inferred an eccentricity of $e_{10}\geq 0.1$ at $10\,\text{Hz}$. However, the evidence for eccentricity was only mild relative to a quasicircular spin-precessing interpretation. Furthermore, these results were obtained using a post-processing likelihood-reweighting technique \cite{Romero_Shaw_2019} rather than by directly sampling the eccentric likelihood.
Independently, the analysis in  Ref.~\cite{gayathri2022eccentricityestimateblackhole} found support for a highly eccentric interpretation, inferring an eccentricity of $0.69^{+0.17}_{-0.22}$ through direct comparisons with numerical relativity (NR) waveforms, although the analysis was limited by the finite set of available NR waveforms. This estimate was subsequently revised to approximately $0.8$ using a $3.5$ post-Newtonian eccentricity estimator in Ref.~\cite{Ciarfella_2022}. More recently, using a similar approach, Ref.~\cite{McMillin2026Disentangling} inferred an eccentricity of $0.38$ at $20\,\text{Hz}$. %, defined at apoastron. 
Using an eccentric spin-precessing waveform model, the analysis in Ref.~\cite{gamboa2026enablinggravitationalwaveastronomyspinprecessing} found weaker support for eccentricity. 
In contrast, several analyses \cite{Iglesias_2024,planas2025timedomainphenomenologicalmultipolarwaveforms,tang2026impacthigherordermodeseccentricity,jxrc-z298,vpyp-nvfp,pompili2026eccentricunboundcompactbinaries}, using eccentric aligned-spin waveform models that include higher-order modes and directly sampling the eccentric likelihood, found no significant evidence for eccentricity. Unbound configurations have also been explored as alternative descriptions of GW190521 \cite{Gamba_2022,lange2026gravitationalwavehyperboliccatalog,pompili2026eccentricunboundcompactbinaries}. Nevertheless, the differing conclusions from eccentric analyses have led to significant tension regarding the eccentric interpretation of GW190521.

Several studies have also investigated spin precession in GW190521 \cite{Biscoveanu_2021,siegel2023ringdowngw190521hintsmultiple,Esteles_2022,PhysRevD.109.024024,Miller_2025}. However, the inferred degree of spin precession can depend on the choice of waveform model. Moreover, features not captured by quasicircular models, particularly those arising from orbital eccentricity, could potentially be interpreted as signatures of spin precession, complicating the interpretation of the inferred precession \cite{Ramos_Buades_2020,Romero_Shaw_2023,Xu_2023,Divyajyoti_2024,divyajyoti2025biasedparameterinferenceeccentric,PhysRevLett.126.201101}. The inferred mass ratio of GW190521 also exhibits model dependence, with some quasicircular spin-precessing phenomenological models favoring asymmetric configurations and producing multimodal mass-ratio posteriors \cite{Nitz_2021,Esteles_2022,Olsen_2021}. Overall, numerous studies have shown that the inferred properties of GW190521 are particularly susceptible to waveform-modeling systematics \cite{Abbott_2020,Esteles_2022,Nitz_2021,48ck-2fff}.

Motivated by the tension among previous eccentricity measurements and the waveform-modeling systematics identified for GW190521, we perform a comprehensive reanalysis of the event using state-of-the-art waveform models, including one that incorporates both eccentricity and spin precession.
We first discuss our analysis settings and the waveform models employed in this work in Sec.~\ref{sec:methods}. We then constrain the eccentricity of GW190521 and assess the impact of the eccentricity prior range on its eccentric interpretation in Sec.~\ref{ssec:gw190521_eccentricity}. In Sec.~\ref{ssec:gw190521_precession}, we investigate spin precession and assess the robustness of its measurement when eccentricity is included in the analysis, followed by a comparison of the different physical hypotheses in Sec.~\ref{ssec:gw190521_hypotheses_comparison}. We constrain the mass parameters in Sec.~\ref{ssec:gw190521_mass}, examining the inferred mass ratio and how the probability of the component masses lying within the PISN mass gap depends on the waveform model and the physical effects included in the analysis. To identify the origin of the observed waveform-modeling systematics, we perform zero-noise injection studies in  Sec.~\ref{sec:bias}. 
Finally, in Sec.~\ref{sec:validation_using_nr}, we perform zero-noise injection--recovery studies with highly eccentric NR waveforms to assess the performance of the eccentric spin-precessing waveform model employed in this work. 

\section{Methods}
\label{sec:methods}
\subsection{Bayesian inference}
We perform a Bayesian analysis of the gravitational-wave strain data for GW190521 \cite{GWOSC_GW190521} from the LIGO Hanford, LIGO Livingston, and Virgo observatories using the {\tt RIFT} parameter-estimation algorithm \cite{PhysRevD.92.023002,lange2018rapidaccurateparameterinference,wofford2023expandingriftimprovingperformance,wagner2025narrowingriftfocusedsimulationbasedinference}. We analyze $4\,$s of data at a sampling rate of $512\,$Hz and evaluate the likelihood over the frequency range $20\text{--}224\,$Hz, matching the configuration provided with the LVK GW190521 parameter-estimation data release \cite{GW190521_data_release}. 
The power spectral densities for all three detectors are taken from the same data release.

For our prior distributions, we use uniform priors on the detector-frame component masses $m_i^\text{det}$, where $i\in \{1,2\}$ labels the two black holes. The spin priors depend on the hypothesis considered. For the spin-precessing hypothesis, we adopt priors uniform in the dimensionless spin magnitudes $\chi_i$ and isotropic in the spin orientations. Under the aligned-spin hypothesis, the spin priors are obtained by projecting a uniform-in-magnitude, isotropic spin distribution onto the direction of the orbital angular momentum \cite{lange2018rapidaccurateparameterinference}. This choice enables a direct comparison between the aligned-spin and spin-precessing analyses. For eccentric analyses, we use uniform priors on the eccentricity and anomaly parameters, using the definitions native to each waveform model. For the binary orientation, we adopt priors that are uniform in the coalescence phase and polarization angle, and in the cosine of the inclination. We use an isotropic prior on the sky location, corresponding to priors uniform in right ascension and in the sine of the declination. The coalescence time is also assigned a uniform prior. Finally, the prior on luminosity distance $D_L$ is taken to be uniform in comoving volume and source-frame time.

When presenting some of our results, we also reformulate the mass and spin parameters in terms of derived parameters commonly used in GW data analysis, including the mass ratio $q=m_2/m_1\leq 1$; the effective aligned-spin parameter $\chi_{\text{eff}}=(m_1\chi_{1\parallel}+m_2\chi_{2\parallel})/(m_1+m_2)$, where $\chi_{i\parallel}$ denotes the component of the dimensionless spin along the orbital angular momentum; the effective precession-spin parameter \cite{Schmidt_2015} $\chi_\text{p}=\max\left(\chi_{1\perp},\frac{4q+3}{4+3q}q\chi_{2\perp}\right)$, where $\chi_{i\perp}$ denotes the spin component in the orbital plane; and the source-frame component masses $m_i^{\rm src}=m_i^{\rm det}/(1+z)$, where $z$ denotes the cosmological redshift inferred from $D_L$ assuming a Planck15 cosmology.

To mitigate spectral leakage when transforming time-domain waveforms into the frequency domain, we taper the beginning of each waveform using the standard tapering routines implemented in the {\tt RIFT} parameter-estimation code. However, for highly eccentric systems, whose waveforms can be substantially shorter than their quasicircular counterparts with the same starting frequency, we find that the standard prescription can result in excessive tapering and consequently bias the inferred $D_L$ and $m_i^\text{src}$. For our high-eccentricity analyses, we therefore modify the tapering prescription. We discuss this further in Appendix~\ref{app:excessive_taper}.

\subsection{Waveform models}
To constrain the orbital eccentricity of this event, we employ three waveform models that incorporate eccentricity: {\tt TEOBResumS-Dal\'{\i}} \cite{PhysRevD.110.024031} (hereafter referred to as {\tt TEOBResumS}), {\tt IMRPhenomTEHM} \cite{Planas_2026} and {\tt SEOBNRv5EHM} \cite{jxrc-z298}. The last two models are restricted to aligned-spin configurations, whereas {\tt TEOBResumS} can generate eccentric waveforms for systems with generic spin orientations. For each waveform model, we measure the eccentricity $e_{5}$ at $5$\,Hz. We choose the $5$\,Hz reference frequency, below the lower frequency cutoff of the likelihood, to measure the eccentricity at an earlier stage of the binary evolution. Furthermore, we report $e_{5}$ using the eccentricity convention native to each waveform model. The short duration of GW190521 prevents us from obtaining an eccentricity estimate using {\tt gw\_eccentricity} \cite{Shaikh_2023,shaikh2025definingeccentricityspinprecessingbinaries}. 

We include all available gravitational-wave modes $h_{\ell,m}$ with $\ell\leq4$ for each waveform model. In the aligned-spin limit ({\tt-E}), {\tt TEOBResumS} includes $(\ell,|m|)=\{(2,2),(2,1),(3,3),(4,4)\}$, as does {\tt IMRPhenomTEHM}, while {\tt SEOBNRv5EHM} includes $\{(2,2),(2,1),(3,3),(3,2),(4,4),(4,3)\}$. For analyses with {\tt TEOBResumS} and {\tt IMRPhenomTEHM}, we perform inference over $e_{5} \in [0.0, 0.9]$, unless stated otherwise. However, for analyses with {\tt SEOBNRv5EHM}, we restrict the prior to $e_{5} \in [0.0, 0.4]$, as we encounter waveform-generation failures at higher eccentricities. We also experienced failures with {\tt TEOBResumS} near the upper limit of $0.9$; however, we do not expect this to impact the inferred distribution, as the posterior support is negligible in that region of parameter space.

To constrain the in-plane spin components of GW190521, and hence the degree of spin precession in the system, we perform inference using the  {\tt TEOBResumS} \cite{PhysRevD.110.024031}, {\tt IMRPhenomTPHM} \cite{Estell_s_2022_imrphenomtphm}, {\tt IMRPhenomXPHM} \cite{PhysRevD.103.104056}, {\tt NRSur7dq4} \cite{Varma_2019} and {\tt SEOBNRv5PHM} \cite{Ramos_Buades_2023} waveform models. For {\tt TEOBResumS}, we consider both the quasicircular spin-precessing ({\tt -P}) and eccentric, spin-precessing ({\tt -E+P}) configurations. The $\ell\leq4$ co-precessing modes included in our analysis are $(\ell,|m|)=\{(2,2),(2,1),(3,3),(4,4)\}$ for {\tt TEOBResumS} and {\tt IMRPhenomTPHM}, $\{(2,2),(2,1),(3,3),(3,2),(4,4)\}$ for {\tt IMRPhenomXPHM}, $\{(2,2),(2,1),(3,3),(3,2),(4,4),(4,3)\}$ for {\tt SEOBNRv5PHM}, and  all $\ell\leq4$ for {\tt NRSur7dq4}. For all waveform models, we set the starting frequency $f_{\rm min}$ to $5$\,Hz and define frequency-dependent quantities at the same frequency.

\section{Parameter inference of GW190521}
\label{sec:PE_GW190521}
In this section, we present the results of our analysis of GW190521. We divide this section into four subsections, each addressing key questions. In Sec.~\ref{ssec:gw190521_eccentricity}, we constrain the orbital eccentricity of the signal and investigate the origin of discrepancies among eccentric interpretations reported in the literature. In Sec.~\ref{ssec:gw190521_precession}, we constrain the spin precession of the system using both eccentric spin-precessing and quasicircular spin-precessing waveform models, and evaluate the impact of including eccentricity on spin-precession measurements.  In Sec.~\ref{ssec:gw190521_hypotheses_comparison}, we investigate which of the three hypotheses (eccentric aligned-spin, quasicircular spin-precessing, and eccentric spin-precessing) is preferred by the data. Finally, in Sec.~\ref{ssec:gw190521_mass}, we compare the mass-ratio and source-frame mass posteriors obtained with different waveform models and assess how the probability of the component masses lying within the PISN mass gap depends on the waveform model and the physical effects included in the analysis.

\subsection{Eccentricity}
\label{ssec:gw190521_eccentricity}
\begin{figure}
    \centering
    \includegraphics[scale=0.52]{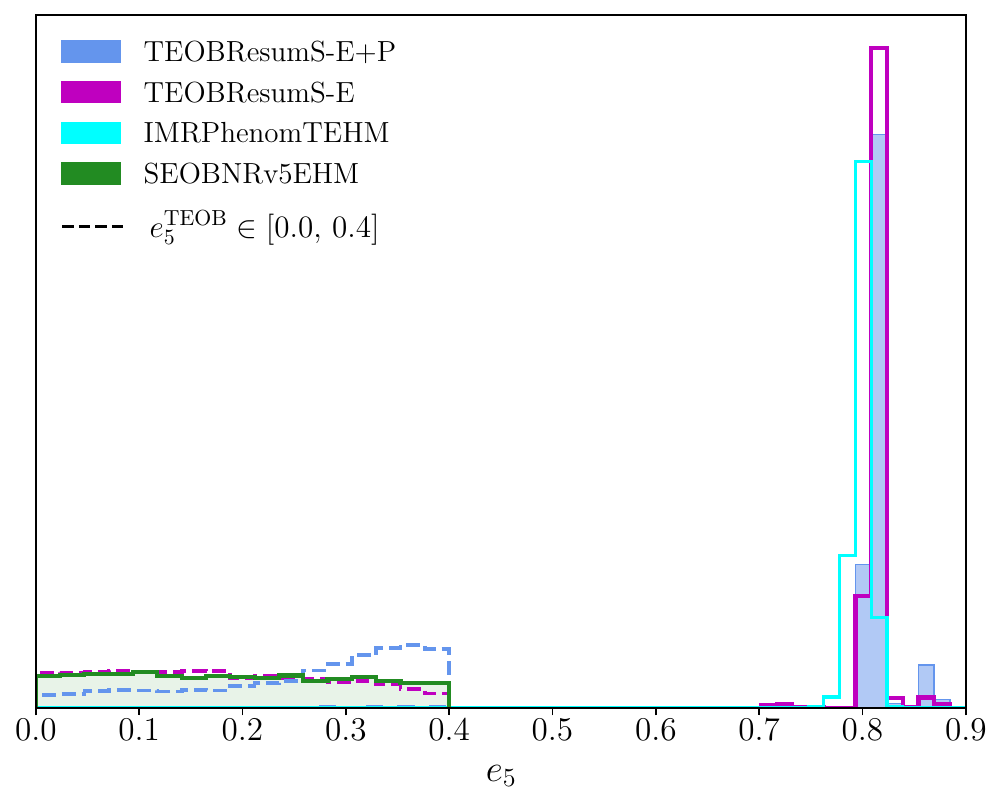}
    \caption{One-dimensional posterior distributions of $e_5$ for GW190521 obtained using {\tt TEOBResumS}, {\tt IMRPhenomTEHM}, and {\tt SEOBNRv5EHM}. Results for {\tt TEOBResumS} are shown under both the eccentric, spin-precessing ({\tt-E+P}) and eccentric, aligned-spin ({\tt-E}) hypotheses. Dashed curves indicate results obtained using  {\tt TEOBResumS} but with a restricted prior $e_5 \in [0.0, 0.4]$.}
    \label{fig:gw190521_eccentricity_comparison}
\end{figure}

The $e_{5}$ posteriors obtained from the analysis of the data are illustrated in Fig.~\ref{fig:gw190521_eccentricity_comparison}, with the corresponding parameter estimates summarized in Table~\ref{tab:gw190521_summary}. From the figure, as well as Table~\ref{tab:gw190521_summary}, we see that {\tt TEOBResumS-E} yields a confident nonzero eccentricity, with $e_{5} = 0.81^{+0.01}_{-0.01}$. This preference for nonzero eccentricity persists  when we analyze the event using {\tt IMRPhenomTEHM}, inferring $e_{5} = 0.80^{+0.01}_{-0.01}$. 
Furthermore, the preference for nonzero eccentricity continues to persist when spin-precession is also included in the analysis.
Analyzing the event using {\tt TEOBResumS-E+P}, we infer $e_{5} = 0.81^{+0.05}_{-0.01}$. Thus, two independent waveform models favor a highly eccentric interpretation of GW190521, and within {\tt TEOBResumS}, this interpretation remains robust when spin precession is included in the analysis.

In contrast, {\tt SEOBNRv5EHM}, for which the analysis is restricted to $e_5\in[0,0.4]$, infers a nearly flat posterior, yielding $e_{5} = 0.19^{+0.19}_{-0.17}$, with significant support for zero eccentricity. However, relative to the {\tt TEOBResumS-E} analysis over the full range, it provides a poorer fit to the data, with a maximum log-likelihood $\ln \mathcal{L}_{\text{max}}$ smaller by a value of $\sim 10$ and log Bayes' factor $\ln \mathrm{BF}$ of $\sim3$, in support of the {\tt TEOBResumS-E} analysis. 

To investigate whether the source of this discrepancy is due to intrinsic differences between the two models or due to the limited $e_5$ prior range in the {\tt SEOBNRv5EHM} analysis, we reanalyze the event using {\tt TEOBResumS-E}, restricting the eccentricity to $e_{5} \in [0.0, 0.4]$. This allows for a direct comparison with the results obtained using {\tt SEOBNRv5EHM}. As shown in Fig.~\ref{fig:gw190521_eccentricity_comparison}, this restricted analysis produces an $e_{5}$ posterior that closely matches the {\tt SEOBNRv5EHM} result. This agreement is also illustrated in the left panel of Fig.~\ref{fig:gw190521_mass_comparison_ecc}, where the source frame component masses obtained from these two analyses are in agreement. This agreement is also found in detector frame masses and spin components.  We obtain a similarly uninformative $e_5$ posterior when restricting the {\tt IMRPhenomTEHM} analysis to $e_5\leq0.4$, although we omit this result for clarity. 

These results demonstrate that the apparent discrepancy between the waveform models is primarily driven by the range of eccentricities accessible to the analyses: when restricted to the same eccentricity range, the three models yield consistent, weakly informative eccentricity posteriors.  To determine the minimum upper prior bound $e_5^{\mathrm{max}}$ required to recover support for nonzero eccentricity, we progressively increase $e_5^{\mathrm{max}}$ from $0.4$ to $0.9$ in increments of $0.1$. We find that $e_5^{\mathrm{max}}$ must extend beyond approximately $0.8$ for the analysis to recover support for nonzero eccentricity. Details of this study are presented in Appendix~\ref{app:gw190521_e5_upper_limit}. This behavior likely arises because the likelihood is not a monotonically increasing function of eccentricity: restricting the prior below the high-likelihood region at large eccentricities therefore does not cause the posterior to rail against the upper prior bound. Instead, a highly eccentric interpretation emerges only once the prior extends into this high-likelihood region. This provides an explanation for why several previous analyses with more restrictive prior ranges on $e$ did not find evidence for eccentricity. Furthermore, the broader analysis yields an increase in  $\ln \mathcal{L}_{\text{max}}$ of $\sim9$ and is favored by a $\ln\mathrm{BF}\sim4.9$, showing the data favor the broader eccentricity range and, consequently, the high-eccentricity solution accessible within it.

\begin{table*}
\centering
% \resizebox{\textwidth}{!}{
\addtolength{\tabcolsep}{3pt} 
\renewcommand{\arraystretch}{1.3}
     \begin{tabular}{| c |c |c |c |c |c |c |c |c |c |}
     \hline
      Model & $m^\text{src}_1$ & $m^\text{src}_2$  & $M^\text{det}_{\text{total}}$  & $q$ & $\chi_{\text{eff}}$ & $\chi_{\text{p}}$ & $e_5$ & $D_L$ & $\ln Z$ ($\ln \mathcal{L}_{\text{max}}$)  \\[0.5ex] 
\hline\hline

{TEOBResumS-E+P} & $98^{+23}_{-18}$ & $72^{+17}_{-14}$ &
$277^{+38}_{-23}$ & $0.74^{+0.23}_{-0.19}$ &
$0.10^{+0.28}_{-0.23}$ & $0.31^{+0.36}_{-0.23}$ &
$0.81^{+0.05}_{-0.01}$\textsuperscript{*} & $3.9^{+1.6}_{-1.7}$ & $82.3~(116.5)$\\

{TEOBResumS-E} & $96^{+23}_{-17}$ & $73^{+17}_{-14}$ &
$281^{+31}_{-23}$ & $0.77^{+0.20}_{-0.20}$ &
$0.18^{+0.21}_{-0.24}$ & ... &
$0.81^{+0.01}_{-0.01}$\textsuperscript{*} & $4.2^{+1.5}_{-1.6}$ & $82.9~(116.4)$\\

{IMRPhenomTEHM} & $101^{+17}_{-18}$ & $76^{+15}_{-14}$ &
$291^{+29}_{-26}$ & $0.75^{+0.20}_{-0.18}$ &
$0.29^{+0.23}_{-0.27}$ & ... &
$0.80^{+0.01}_{-0.01}$ & $4.0^{+1.6}_{-1.5}$ & $84.4~(115.9)$\\

{SEOBNRv5EHM} & $86^{+17}_{-14}$ & $61^{+13}_{-16}$ &
$259^{+27}_{-25}$ & $0.71^{+0.24}_{-0.26}$ &
$0.04^{+0.26}_{-0.27}$ & ... &
$0.19^{+0.19}_{-0.17}$$^\dagger$ & $4.9^{+1.6}_{-1.5}$ & $79.4~(106.0)$\\

{TEOBResumS-P} & $98^{+19}_{-17}$ & $70^{+20}_{-18}$ &
$264^{+36}_{-34}$ & $0.72^{+0.22}_{-0.20}$ &
$0.00^{+0.32}_{-0.49}$\textsuperscript{*} & $0.56^{+0.32}_{-0.36}$ &
$...$ & $3.3^{+2.2}_{-1.6}$ & $80.3~(113.4)$\\

{IMRPhenomTPHM} & $98^{+18}_{-17}$ & $70^{+20}_{-18}$ &
$260^{+33}_{-31}$ & $0.72^{+0.23}_{-0.21}$ &
$-0.07^{+0.32}_{-0.33}$ & $0.59^{+0.32}_{-0.38}$ &
$...$ & $3.2^{+2.0}_{-1.4}$ & $80.3~(112.6)$\\

{IMRPhenomXPHM} & $95^{+21}_{-18}$ & $56^{+16}_{-19}$\textsuperscript{*} &
$239^{+42}_{-31}$ & $0.58^{+0.32}_{-0.24}$\textsuperscript{*} &
$-0.20^{+0.42}_{-0.38}$\textsuperscript{*} & $0.42^{+0.37}_{-0.31}$ &
$...$ & $3.5^{+2.5}_{-1.7}$ & $80.3~(114.5)$\\

{NRSur7dq4} & $86^{+21}_{-14}$ & $67^{+20}_{-18}$ &
$268^{+27}_{-27}$ & $0.81^{+0.17}_{-0.29}$ &
$0.02^{+0.28}_{-0.33}$ & $0.58^{+0.33}_{-0.41}$ &
$...$ & $4.8^{+2.2}_{-2.5}$ & $81.7~(114.0)$\\

{SEOBNRv5PHM} & $93^{+19}_{-17}$ & $66^{+19}_{-18}$ &
$268^{+32}_{-29}$ & $0.72^{+0.24}_{-0.24}$ &
$0.04^{+0.28}_{-0.31}$ & $0.56^{+0.35}_{-0.41}$ &
$...$ & $4.3^{+1.9}_{-2.0}$ & $81.2~(111.9)$\\

% \hline

% {GW231123} & $134^{+19}_{-14}$ & $93^{+14}_{-11}$ &
% $365^{+18}_{-23}$ & $0.69^{+0.12}_{-0.09}$ &
% $0.43^{+0.08}_{-0.09}$ & $0.71^{+0.12}_{-0.15}$ &
% $0.06^{+0.08}_{-0.06}$ & $3.6^{+1.0}_{-1.2}$ & ...\\

\hline
     \end{tabular}
     \caption{Summary of median values and $90\%$ equal-tailed credible intervals for the source properties of GW190521 inferred using multiple waveform models. We also report the logarithm of the Bayesian evidence, $\ln Z$, together with the maximum log-likelihood, $\ln \mathcal{L}_{\mathrm{max}}$, for each analysis. Masses are given in $M_\odot$ and luminosity distance $D_L$ is reported in Gpc. An asterisk ($*$) denotes a multimodal posterior, while a dagger ($\dagger$) denotes an analysis performed with the reduced $e_5$ range.}
    \label{tab:gw190521_summary}
\end{table*}

 \subsection{Spin precession}
\label{ssec:gw190521_precession}

\begin{figure}
    \centering
    \includegraphics[scale=0.52]{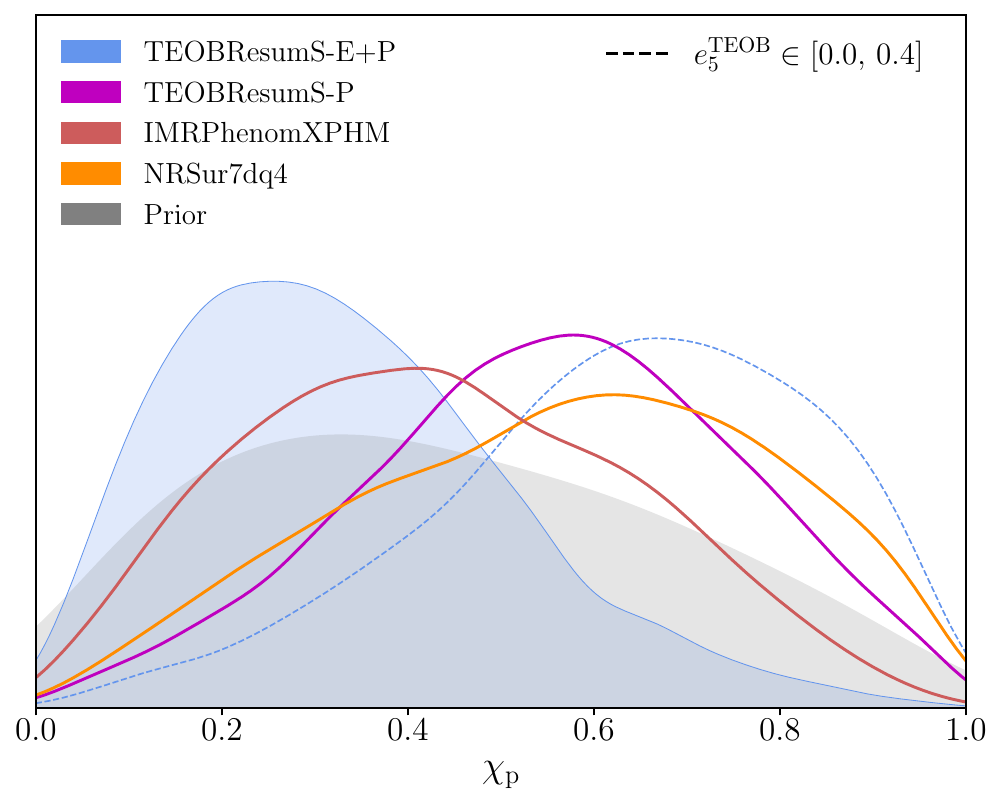}
    \caption{One-dimensional posterior distributions of $\chi_\text{p}$ for GW190521 obtained using {\tt TEOBResumS-E+P},  {\tt TEOBResumS-P},  {\tt IMRPhenomXPHM} and {\tt NRSur7dq4}. Dashed curve corresponds to the {\tt TEOBResumS-E+P} analysis with a restricted prior $e_5 \in [0.0, 0.4]$. The prior distribution for $\chi_\mathrm{p}$ is also shown.}
    \label{fig:gw190521_precession_comparison}
\end{figure}

The $\chi_\text{p}$ posteriors obtained from the analysis of the data using select waveform models are shown in Fig.~\ref{fig:gw190521_precession_comparison}, while the parameter estimates obtained using all waveform models employed in this work are summarized in Table~\ref{tab:gw190521_summary}. As shown in the table, {\tt IMRPhenomTPHM}, {\tt NRSur7dq4}, and {\tt SEOBNRv5PHM} recover similar $\chi_\text{p}$ posteriors. For clarity, we therefore omit the {\tt IMRPhenomTPHM} and {\tt SEOBNRv5PHM} posteriors from Fig.~\ref{fig:gw190521_precession_comparison}. {\tt TEOBResumS-P} also yields a broadly similar posterior, although with a slightly sharper peak. These four models infer median values of $\chi_\text{p}\sim0.6$, whereas {\tt IMRPhenomXPHM} yields a somewhat different posterior with a lower median of $\chi_\text{p}\sim0.45$. Despite these differences, all quasicircular analyses favor some degree of spin precession. The corresponding maximum log likelihoods are comparable, with  $\ln \mathcal{L}_{\max} \sim113$--$114$ for most models and a slightly lower value of $\sim111$ for {\tt SEOBNRv5PHM}. Furthermore, the $\ln\mathrm{BF}$ values do not indicate a preference for any one of the waveform models considered in the quasicircular analyses.

When eccentricity is included through {\tt TEOBResumS-E+P}, the inferred $\chi_\mathrm{p}$ posterior shifts toward lower values relative to the quasicircular analyses. This suggests that part of the support for large spin precession in the quasicircular analyses arises from those models attempting to accommodate eccentric features present in the data. 
Simultaneously,  $\ln \mathcal{L}_{\max}$ increases from $113.4$ to $116.5$, indicating that the inclusion of eccentricity improves the agreement between the model and the data. The increase in likelihood suggests that eccentricity captures waveform structure that is not fully accounted for in the quasicircular analyses. 
For completeness, we also show the posterior obtained when the eccentricity prior is restricted to $e_5 \in [0,0.4]$. In this case, the $\chi_\mathrm{p}$ posterior shifts toward larger values, indicating that restricting the allowed eccentricity forces the inference to accommodate features in the data through increased spin precession instead. 
However, the restricted-eccentricity analysis yields a lower maximum likelihood and Bayesian evidence, with $\Delta \ln \mathcal{L}_{\max}\sim3.2$ and $\ln \text{BF}\sim1.7$ in favor of the full eccentricity range.

Overall, we find that the inferred degree of spin precession in GW190521 depends on the waveform model and substantially depends on the level of physics included in the analysis. 
While quasicircular spin-precessing models generally favor moderate-to-large $\chi_\text{p}$ values, the inclusion of eccentricity over the full range shifts the inference toward smaller $\chi_\text{p}$ values. This behavior is illustrated most clearly by comparing {\tt TEOBResumS-P} and {\tt TEOBResumS-E+P}, which differ primarily through the inclusion of eccentricity. Taken together, these results provide evidence for a degeneracy between eccentricity and spin precession for GW190521, and that spin precession in GW190521 is not robust to the inclusion of eccentricity.

\subsection{Eccentric or spin precessing}
\label{ssec:gw190521_hypotheses_comparison}
To assess which hypothesis is favored by the data, we first compare the analyses performed within the {\tt TEOBResumS} framework. The {\tt TEOBResumS-E+P} and {\tt TEOBResumS-E} analyses achieve nearly identical maximum likelihoods, with $\ln\mathcal{L}_\text{max}\sim116$, indicating that the inclusion of spin precession does not significantly improve the fit to the data once eccentricity is included. Furthermore, the $\ln\mathrm{BF}$ shows no preference between the eccentric aligned-spin and eccentric spin-precessing hypotheses. In contrast, comparing {\tt TEOBResumS-E} with {\tt TEOBResumS-P}, we find that the eccentric aligned-spin hypothesis provides a better fit to the data, with an increase in $\ln\mathcal{L}_{\max}$ of $\sim3$ and a $\ln\mathrm{BF}\sim2.6$ in its favor. Thus, within the {\tt TEOBResumS} framework, the data favor the eccentric aligned-spin hypothesis (eccentric spin-precessing) over the quasicircular spin-precessing hypothesis. 

We find a similar preference within the {\tt IMRPhenomT} framework: {\tt IMRPhenomTEHM} achieves a $\ln\mathcal{L}_{\max}$ larger by $\sim3$ than {\tt IMRPhenomTPHM}, with a $\ln\text{BF}\sim4$ in its favor. Therefore, across two independent waveform frameworks, the data favor the eccentric aligned-spin hypothesis over the quasicircular spin-precessing hypothesis.

\subsection{Mass parameters}
\label{ssec:gw190521_mass}
\begin{figure}
    \centering
    \includegraphics[scale=0.52]{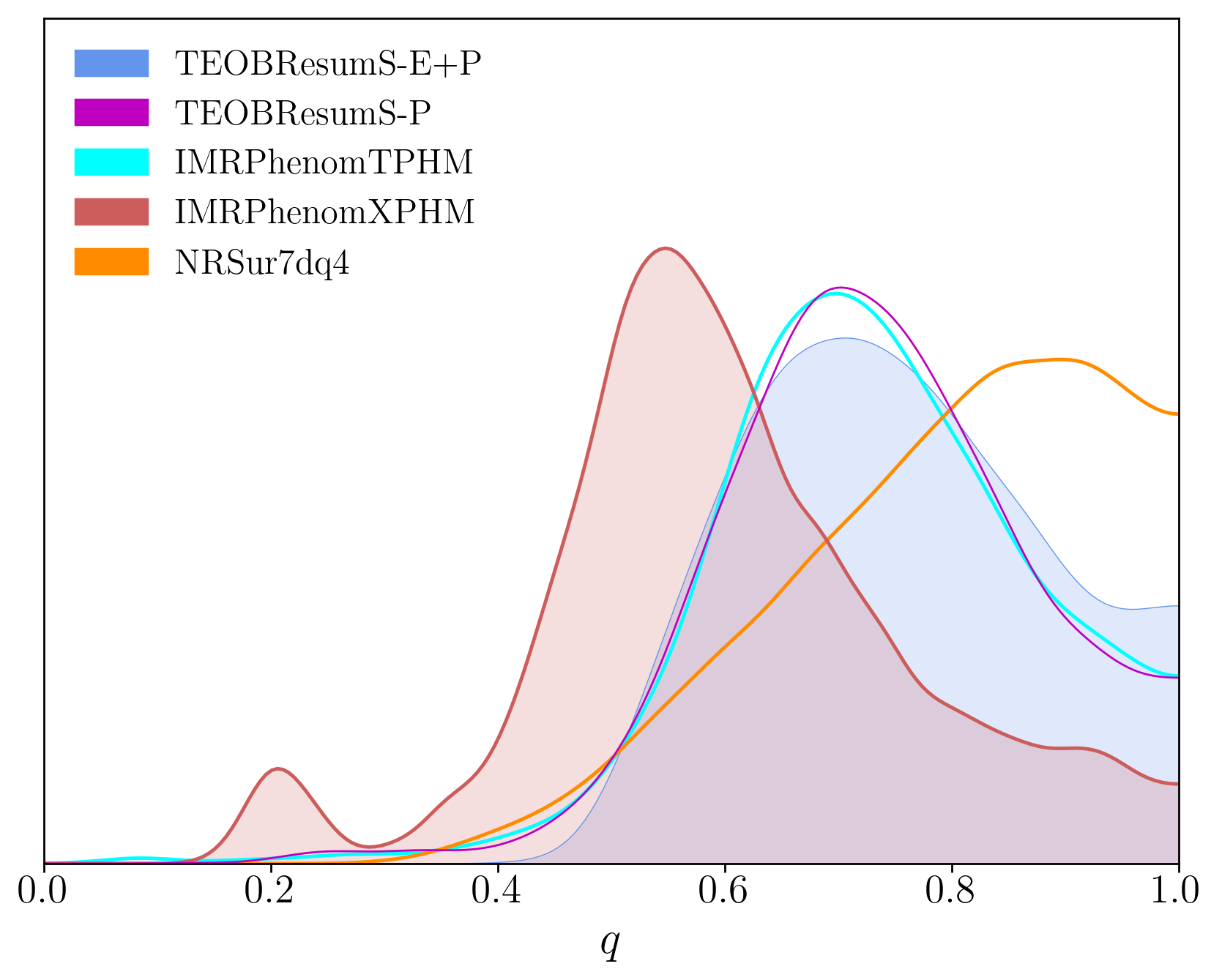}
    \caption{One-dimensional posterior distributions of $q$ for GW190521 obtained using  {\tt TEOBResumS-E+P},  {\tt TEOBResumS-P}, {\tt IMRPhenomTPHM}, {\tt IMRPhenomXPHM} and {\tt NRSur7dq4}. All models predominantly favor a comparable-mass interpretation, with {\tt IMRPhenomXPHM} exhibiting a subdominant mode at more asymmetric mass ratios.}
    \label{fig:gw190521_q_comparison}
\end{figure}

\begin{figure*}
    \centering
    \includegraphics[scale=0.52]{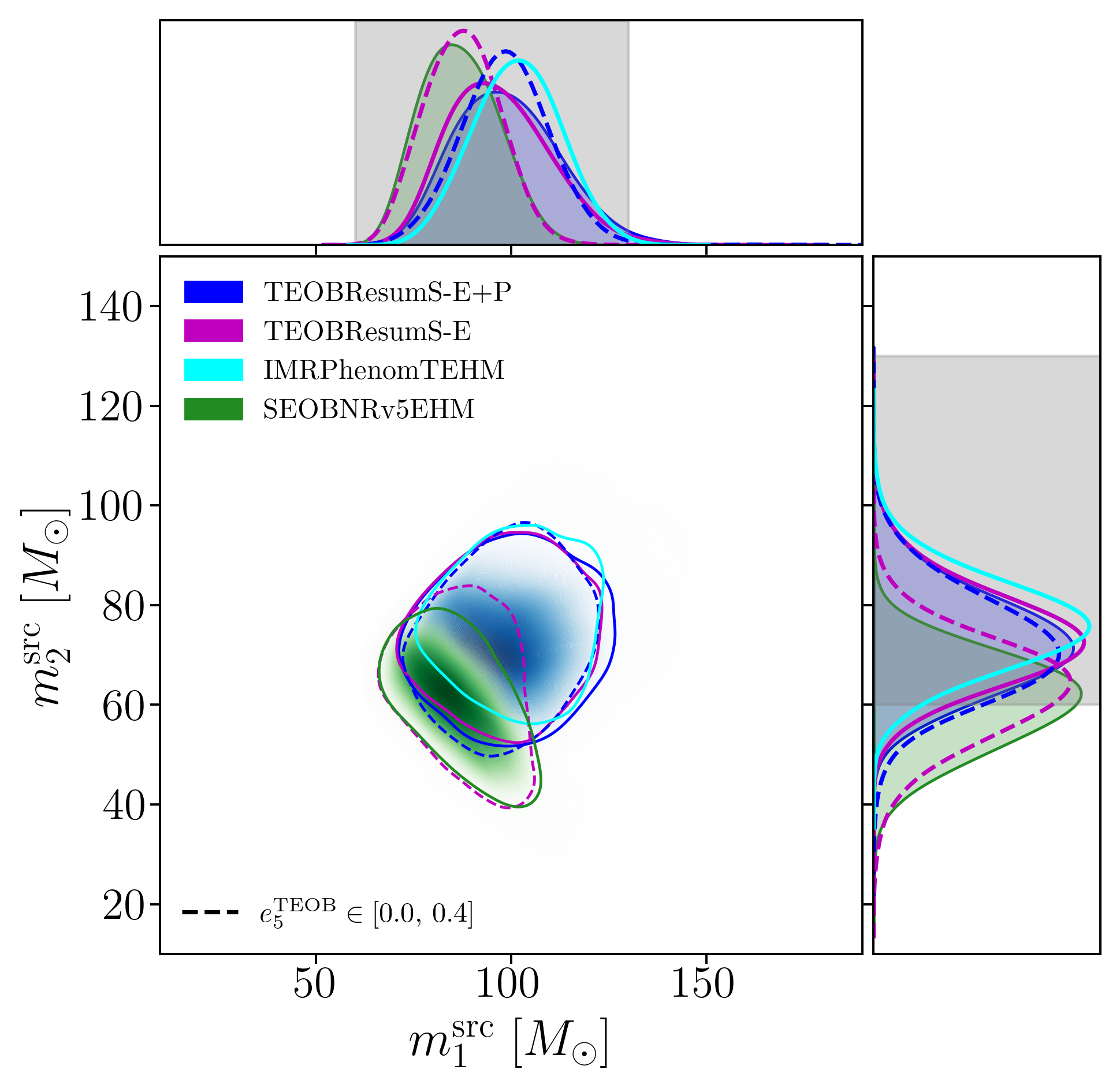}
    \includegraphics[scale=0.52]{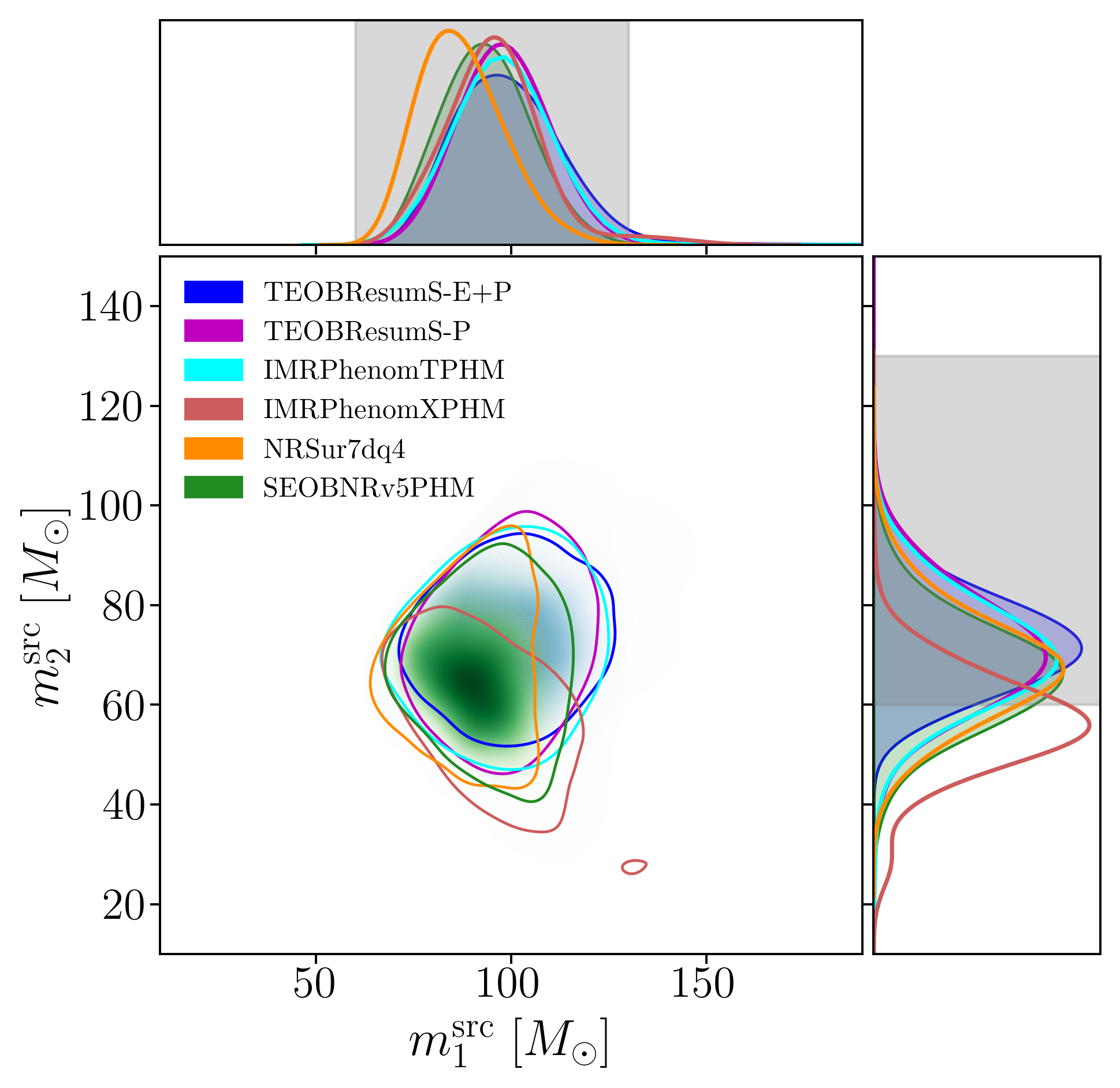}
    \caption{One- and two-dimensional posterior distributions of source-frame masses for GW190521 obtained using models including eccentricity (\textit{left panel}) and spin precession (\textit{right panel}). Results for {\tt TEOBResumS} are shown under eccentric, spin-precessing ({\tt -E+P}), eccentric, aligned-spin ({\tt -E}), and quasicircular spin-precessing ({\tt -P}) hypotheses. Solid curves use the full prior range, while dashed curves correspond to a restricted prior $e_5 \in [0.0, 0.4]$. The shaded gray region denotes the PISN mass gap ($\sim 60\text{--}130\,M_\odot$).}
    \label{fig:gw190521_mass_comparison_ecc}
\end{figure*}
The mass-ratio estimates and their associated uncertainties obtained with all waveform models employed in this work are summarized in Table~\ref{tab:gw190521_summary}, with posterior distributions for selected models shown in Fig.~\ref{fig:gw190521_q_comparison}. For clarity, we show posteriors from {\tt TEOBResumS-E+P}, {\tt TEOBResumS-P}, {\tt IMRPhenomTPHM}, {\tt IMRPhenomXPHM}, and {\tt NRSur7dq4}. All waveform models favor a comparable-mass binary, although there are some differences in the inferred distributions. In particular, {\tt IMRPhenomXPHM} yields a bimodal posterior, with a secondary mode near $q\sim0.2$ and support extending down to $q\sim0.1$. This secondary mode corresponds to a substantially heavier primary black hole, visible as a small isolated region in the right panel of Fig.~\ref{fig:gw190521_mass_comparison_ecc}, and a lighter secondary black hole that lies below the PISN gap. Interestingly, the $\ln \mathcal{L}_\text{max}$ point obtained with {\tt IMRPhenomXPHM} lies within this secondary mode. Furthermore, the {\tt IMRPhenomTPHM} analysis also exhibits some support for more asymmetric configurations. Although the bulk of its posterior is concentrated at $q\gtrsim0.4$, a tail extends toward $q\sim0.1$. Overall, these results favor a comparable-mass interpretation of GW190521, while indicating that the degree of support for asymmetric configurations is sensitive to the choice of waveform model.

We next examine how the inferred source-frame masses depend on the eccentricity prior. As shown in the left panel of Fig.~\ref{fig:gw190521_mass_comparison_ecc}, restricting the eccentricity range has a pronounced effect on the masses inferred with {\tt TEOBResumS-E}, shifting the component-mass posteriors toward lower values. Consequently, the probability that the secondary lies within the PISN gap decreases from $94.0\%$ for the full eccentricity range to $63.4\%$ for the reduced range, while the corresponding probability for the primary remains high, changing from $99.0\%$ to $99.9\%$. The shift in the component masses is also reflected in the total source-frame mass, which increases from $149^{+22}_{-14}\,M\odot$ for the reduced eccentricity range to $168^{+34}_{-19}\,M\odot$ for the full range. Thus, allowing the analysis to access the high-eccentricity region for the aligned-spin configurations substantially strengthens the support for both components of GW190521 residing within the PISN gap. In contrast, the source-frame mass posteriors obtained with {\tt TEOBResumS-E+P} are considerably less sensitive to the eccentricity prior, resulting in only modest changes in the corresponding PISN-gap probabilities. 

Finally, we compare the source-frame masses inferred using waveform models that include spin precession, with the corresponding posteriors shown in the right panel of Fig.~\ref{fig:gw190521_mass_comparison_ecc}. The $m_1^\mathrm{src}$ posteriors are broadly consistent across models, with {\tt NRSur7dq4} favoring slightly lower masses than the other analyses. Accordingly, all models infer a high probability, $\gtrsim97\%$, that the primary lies within the PISN gap. The inference for the secondary is more sensitive to the waveform model. {\tt TEOBResumS-E+P} yields the largest probability of the secondary lying within the PISN gap, $\sim92\%$, followed by {\tt TEOBResumS-P} and {\tt IMRPhenomTPHM} at $\sim84\%$. {\tt NRSur7dq4} and {\tt SEOBNRv5PHM} yield probabilities of $\sim72$--$76\%$, whereas {\tt IMRPhenomXPHM} gives a substantially lower probability of $\sim33\%$. The latter also supports a secondary mode in which the secondary black hole lies below the PISN gap and the primary lies in the intermediate-mass black-hole regime. Thus, while the conclusion that the primary resides within the PISN gap is robust across the waveform models considered here, the corresponding conclusion for the secondary is substantially more sensitive to waveform-modeling systematics.

\section{Biases from unmodeled eccentricity: Zero-noise injection study}
\label{sec:bias}
To identify the origin of the observed waveform-modeling systematics for GW190521 and to gain further insights into the event, we perform two zero-noise injection studies. In the first study, we inject the $\ln \mathcal{L}_\text{max}$ point obtained from the GW190521 analysis using {\tt TEOBResumS-E}. This configuration has $e_5 = 0.81$, and yields one of the largest $\ln \mathcal{L}_{\max}$ among the analyses considered in Table~\ref{tab:gw190521_summary}. Although the {\tt -E+P} analysis attains a marginally higher maximum likelihood, we choose the {\tt -E} configuration to isolate the impact of eccentricity on the posteriors obtained with quasicircular models. 
We then recover this synthetic signal with a range of waveform models to address the following questions:

\begin{enumerate}
\item Does restricting the eccentricity prior range produce uninformative or weakly constrained $e_5$ posteriors, even when the injected signal is highly eccentric?
\item Can quasicircular spin-precessing waveform models recover significant spin precession when analyzing this highly eccentric non-precessing signal, thereby demonstrating the eccentricity--spin-precession degeneracy?
\item Can the waveform systematics observed in the real event with quasicircular spin-precessing waveform models be explained by model-dependent interactions with the eccentric features of the signal?
\end{enumerate}
\begin{figure*}
    \centering
    \includegraphics[scale=0.4]{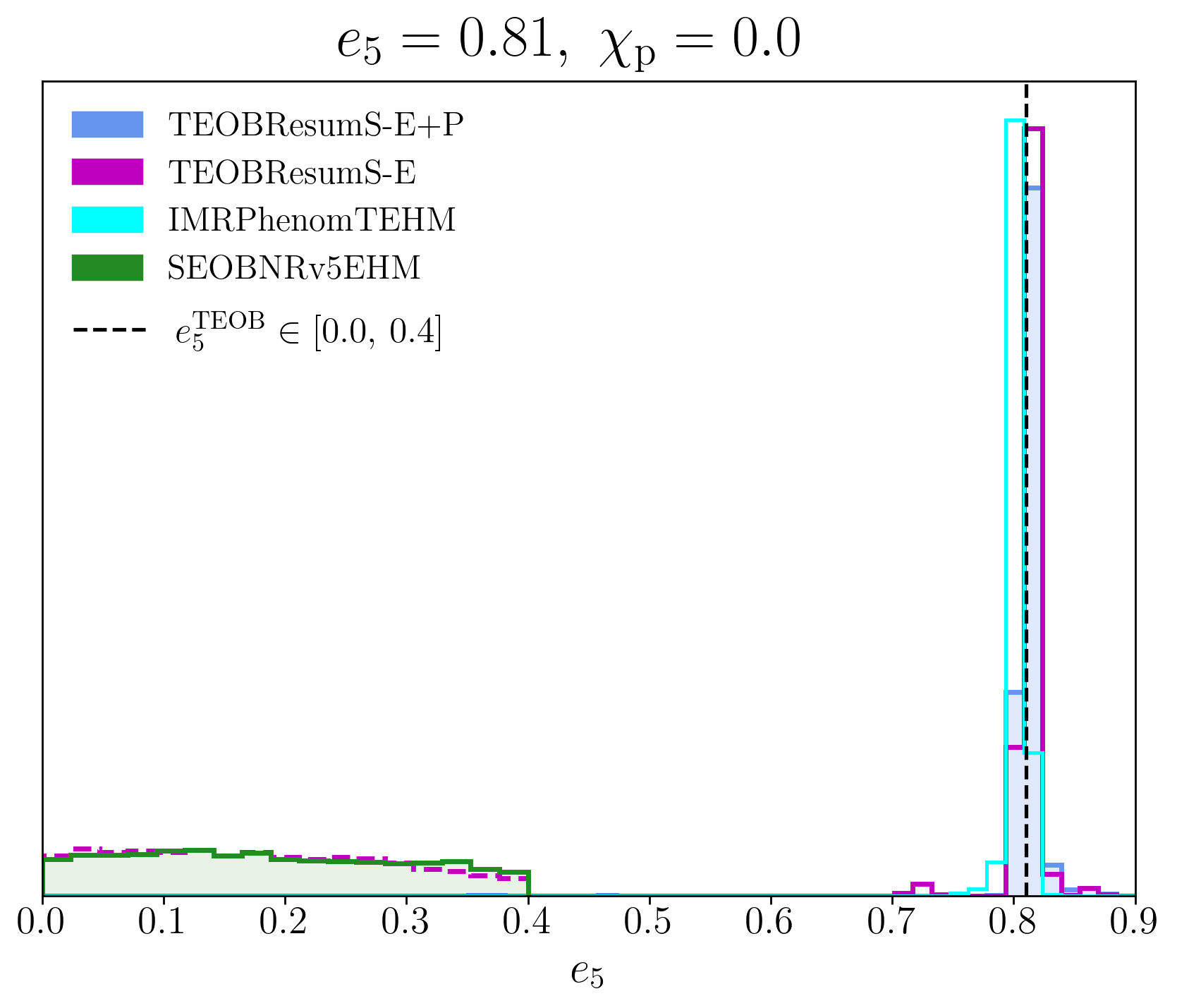}
    \includegraphics[scale=0.4]{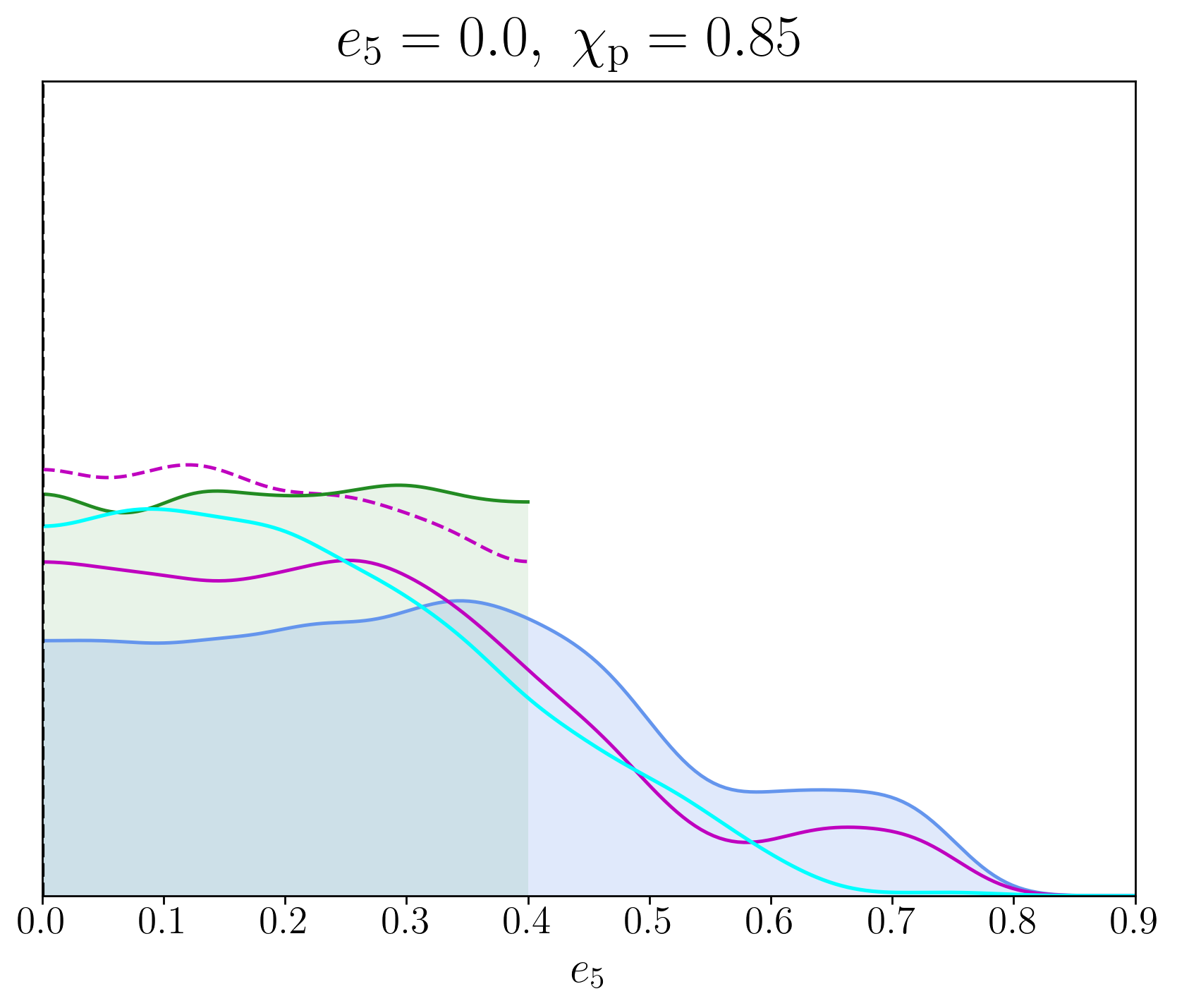}
    \includegraphics[scale=0.4]{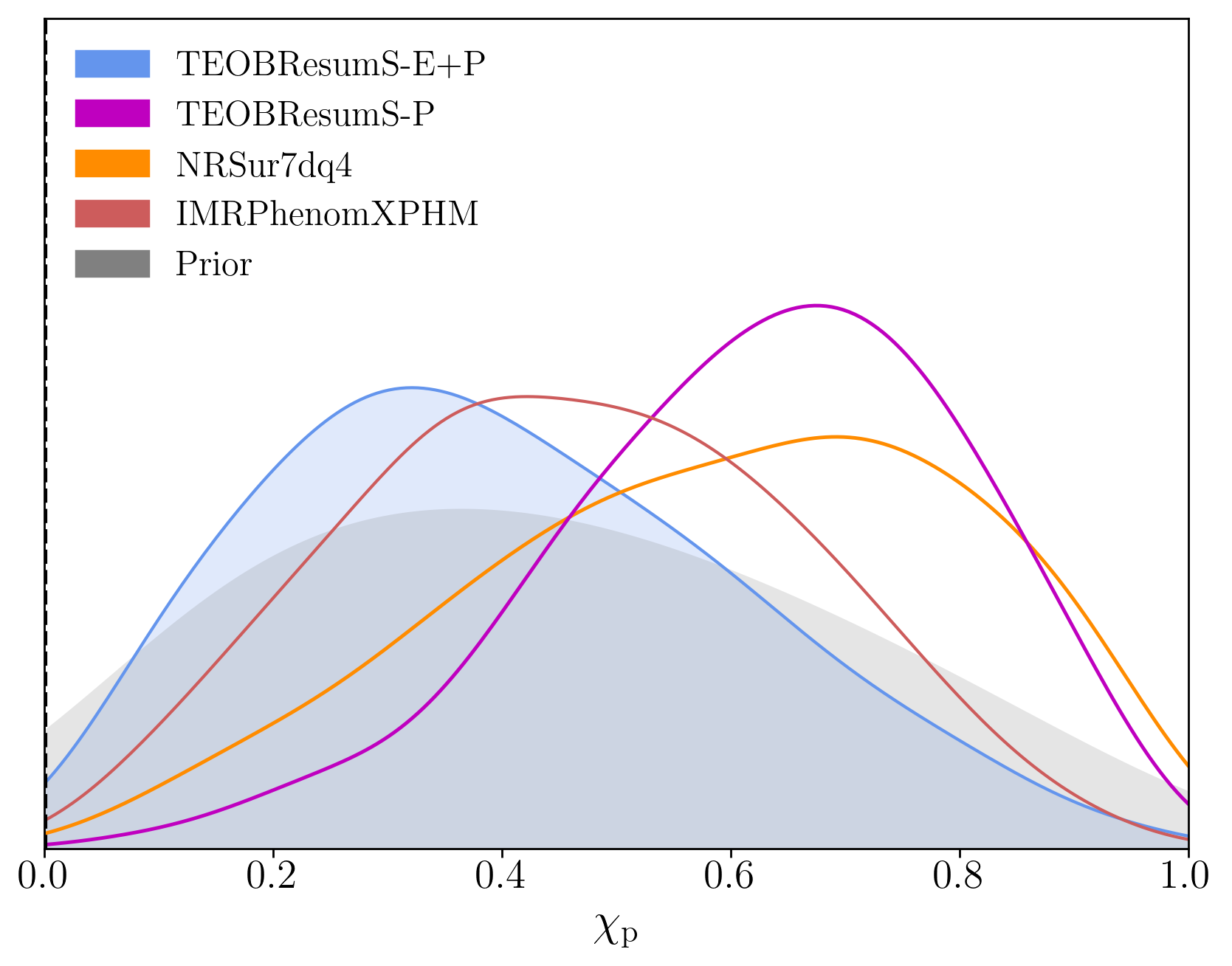}
    \includegraphics[scale=0.4]{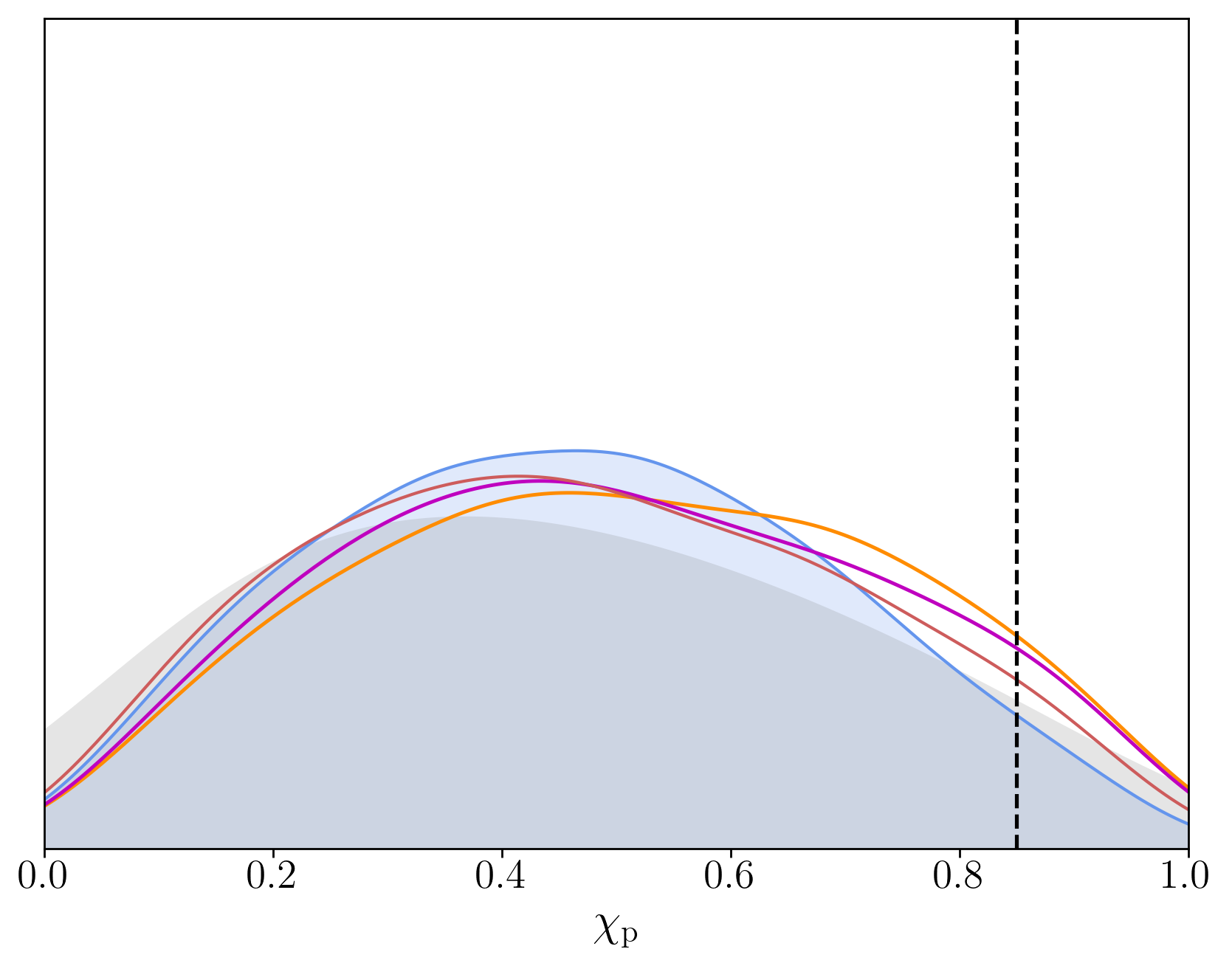}
    \includegraphics[scale=0.4]{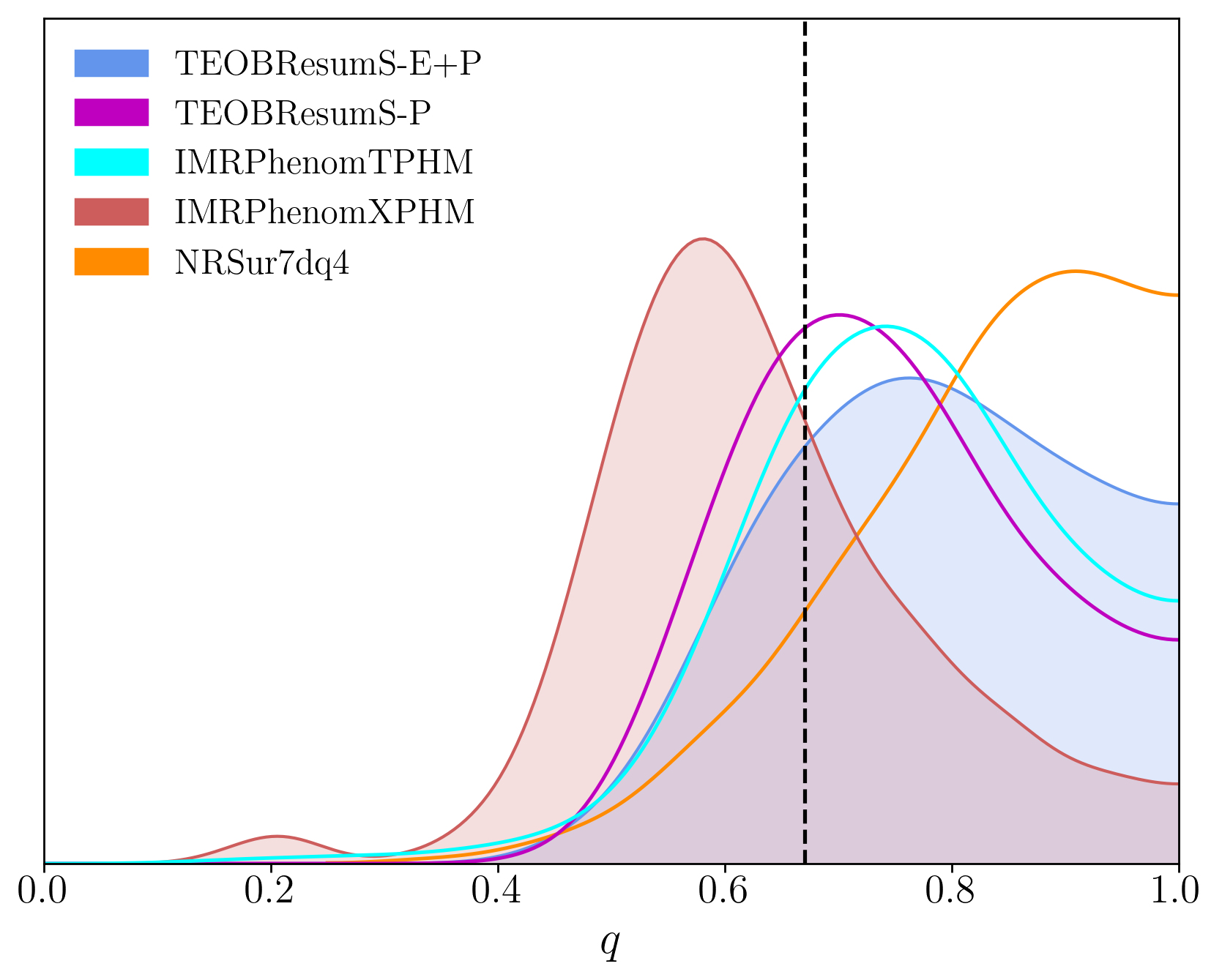}
    \includegraphics[scale=0.4]{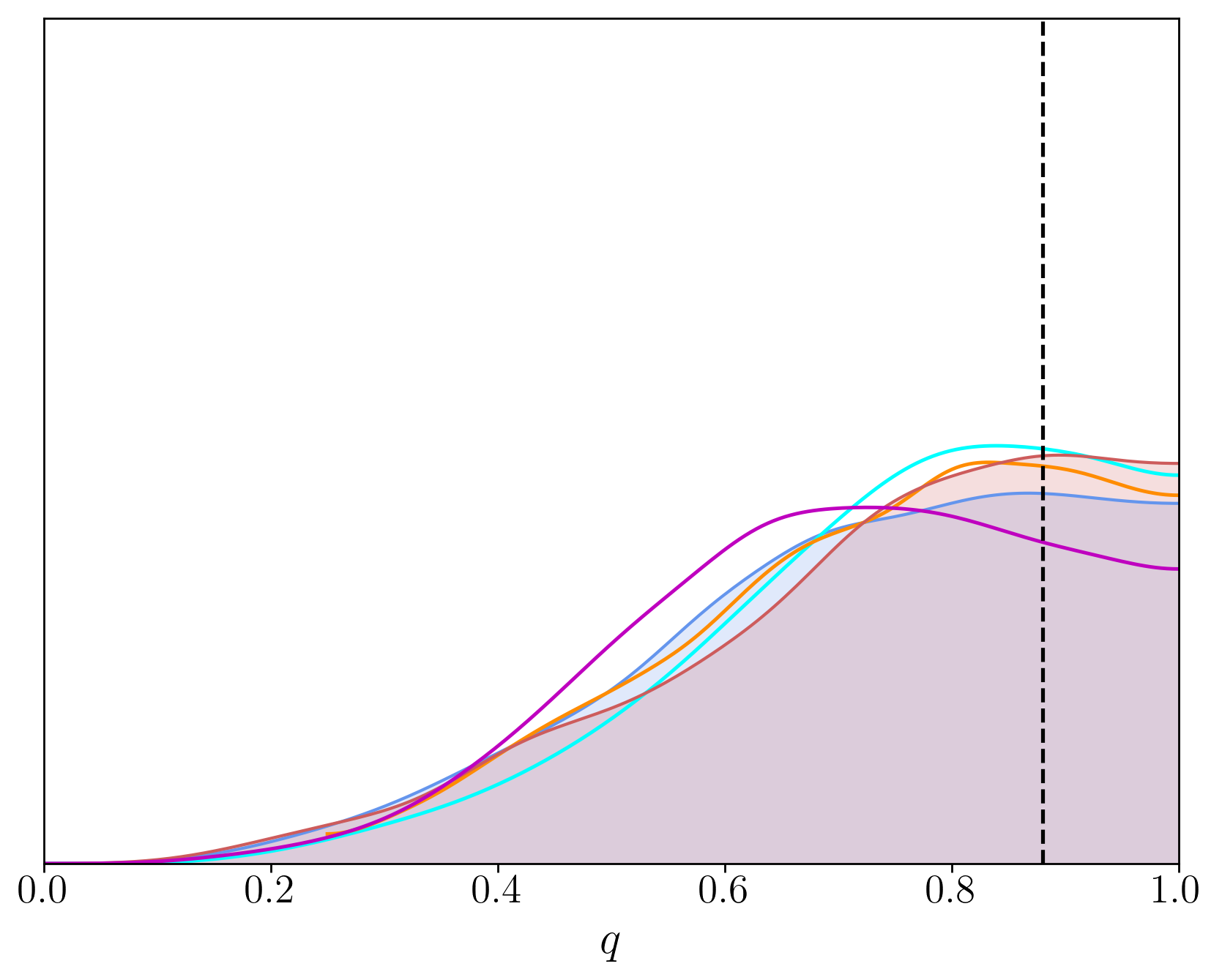}
    \caption{Posterior distributions obtained from the analysis of synthetic data generated using \texttt{TEOBResumS-E} with $e_5 = 0.81$ (left column) and \texttt{NRSur7dq4} with $\chi_\mathrm{p}=0.85$ (right column). The injections were performed at the corresponding $\ln \mathcal{L}_\text{max}$ parameters obtained from the analysis of GW190521 with each waveform model. The eccentric injection exhibits significant waveform-model systematics, while the quasicircular, strongly precessing injection yields consistent results across models.}
    \label{fig:gw190521_max_lnl_comparison}
\end{figure*}

The results of analyzing this synthetic injection are shown in the left column of Fig.~\ref{fig:gw190521_max_lnl_comparison}. We find that when the synthetic signal is analyzed with {\tt TEOBResumS-E+P}, {\tt TEOBResumS-E} and {\tt IMRPhenomTEHM} over the full $e_5$ prior range, the injected eccentricity is successfully recovered. Moreover, the three analyses also achieve comparable maximum likelihoods and Bayesian evidences. However, the two {\tt TEOBResumS}  analyses exhibit mild multimodal structure similar to that observed in the analysis of the real data. These results demonstrate that the injected eccentricity can be successfully recovered by the eccentric models employed in this study and, importantly, that independent waveform models yield broadly consistent eccentricity estimates for a highly eccentric GW190521-like injection. 
In contrast, when the eccentricity prior is restricted to $e_5 \in [0,0.4]$, both {\tt TEOBResumS-E} and {\tt SEOBNRv5EHM} recover nearly flat, weakly constrained eccentricity posteriors. 

Comparing the results obtained under the aligned-spin eccentric hypothesis, we find that restricting the eccentricity range also affects the inferred detector-frame component masses, shifting the recovered posteriors relative to the full-range analysis. We also observe substantial differences in the $\chi_{1z}$ posteriors, whereas the $\chi_{2z}$ posteriors exhibit significant overlap. For the source-frame masses, restricting the eccentricity prior shifts the recovered posteriors toward lower values, whereas allowing the full eccentricity range favors relatively larger masses. Consequently, the estimated probability that the component masses lie within the PISN gap also depends on the adopted eccentricity prior, particularly for the secondary component. Similarly, when comparing the restricted- and full-eccentricity-range analyses under the eccentric spin-precessing hypothesis, we find differences in the detector-frame masses and $\chi_{1z}$ posteriors, while the source-frame mass posteriors show excellent agreement. These trends closely mirror those observed in the analysis of the real GW190521 data. Together, these results demonstrate that, for GW190521-like signals with large eccentricity, restricting the eccentricity prior can obscure the underlying eccentric nature of the source and introduce systematic shifts in the inferred parameters.

Examining the $\chi_\mathrm{p}$ posteriors in the middle-left panel of Fig.~\ref{fig:gw190521_max_lnl_comparison}, we find that quasicircular spin-precessing waveform models infer posteriors that deviate from the prior and show substantial support for significant spin precession despite the injected signal being non-precessing. In particular, {\tt NRSur7dq4} and {\tt TEOBResumS-P} recover posteriors peaking near $\chi_\mathrm{p}\sim0.7$, while {\tt IMRPhenomXPHM} peaks near $\chi_\mathrm{p}\sim0.4$. 
In contrast, {\tt TEOBResumS-E+P} correctly recovers a posterior favoring low $\chi_\text{p}$ values, consistent with the injected non-precessing signal. These results demonstrate that, for GW190521-like systems, analyzing a highly eccentric signal with quasicircular spin-precessing waveform models can lead to spurious evidence for spin precession. This behavior indicates that spin precession can partially mimic unmodeled eccentric features in short, high-mass signals such as GW190521.

The bottom-left panel of Fig.~\ref{fig:gw190521_max_lnl_comparison} shows the recovered $q$ posteriors. While all waveform models favor a relatively comparable-mass system, {\tt IMRPhenomXPHM} exhibits the largest disagreement with the other waveform models, as was also observed in the analysis of the real GW190521 data. In particular, it produces a bimodal posterior with a secondary mode near $q\sim0.2$. Examining the corresponding source-frame mass posteriors, we find that {\tt IMRPhenomXPHM} also favors a smaller secondary mass than the other waveform models. The low-$q$ mode corresponds to a region of parameter space with a substantially heavier primary black hole and a lighter secondary component, again mirroring the behavior observed in the analysis of the real data.

More generally, when analyzing this injection using quasicircular models, we observe varying levels of waveform-model systematics across several inferred parameters. The differences among the posteriors obtained with quasicircular spin-precessing waveform models indicate that different waveform models absorb the missing eccentric features in the signal differently, leading to model-dependent inference. These results suggest that at least part of the waveform-model systematics observed in the real GW190521 analysis arise from neglecting eccentricity in the waveform modeling.

While the preceding injection study strongly suggests that unmodeled eccentricity can explain waveform systematics observed when analyzing GW190521 using quasicircular spin-precessing models, a few questions remain:
\begin{enumerate}
\item Could \texttt{TEOBResumS-E+P} itself incorrectly interpret spin-precession features as eccentricity due to mismodeling?
\item Do quasicircular waveform models exhibit comparable disagreements when analyzing a genuinely quasicircular, strongly precessing signal?
\end{enumerate}
To address these questions, we inject the maximum-likelihood configuration obtained from the {\tt NRSur7dq4} analysis of GW190521. This injection is characterized by $\chi_\mathrm{p}=0.85$. The results are shown in the right column of Fig.~\ref{fig:gw190521_max_lnl_comparison}. We find that {\tt TEOBResumS-E+P}, {\tt TEOBResumS-E} and {\tt IMRPhenomTEHM} recover broad eccentricity posteriors that retain substantial support near $e_5=0$, consistent with the quasicircular nature of the injected signal. Restricting the eccentricity prior again produces broad, weakly informative posteriors.

In contrast to the eccentric injection, the middle-right panel shows that all waveform models recover consistent $\chi_\mathrm{p}$ posteriors despite the large injected spin precession. Similarly, the bottom-right panel shows agreement in the recovered mass-ratio posteriors, with only mild deviations in the {\tt TEOBResumS-P} results. More generally, we find excellent agreement across the inferred intrinsic parameters for both the quasicircular spin-precessing and eccentric, spin-precessing waveform models. These results indicate that {\tt TEOBResumS-E+P} is capable of distinguishing between eccentric and spin-precession features in the signal, even when the precessing morphology is generated using an independent waveform model. 
More importantly, the absence of large waveform-model disagreements in this injection study indicates that the waveform models behave consistently for a genuinely spin-precessing GW190521-like event.

Combined, the two injection studies suggest that many of the waveform-model systematics observed in GW190521 can be understood as different waveform models responding differently to unmodeled eccentric features present in the data.
\section{Validation of Eccentric Inference: Numerical-Relativity Injections}
\label{sec:validation_using_nr}
\begin{table*}
\centering
\renewcommand{\arraystretch}{1.4}
\addtolength{\tabcolsep}{1pt} 
     \begin{tabular}{| c |c | c | c | c | c | c | c | c | c |}
     \hline
      NR waveform & $m^\text{src}_1$ & $m^\text{src}_2$  & $M^\text{det}_{\text{total}}$  & $q$ & $\chi_{\text{eff}}$ & $\chi_{\text{p}}$ & $e_0$ & $D_L$ &  $\ln \mathcal{L}_{\text{max}}$  \\[0.5ex] 
\hline\hline
    {\tt RIT\_eBBH\_1637} & $80$ & $80$ & $290$ & $1$ & $0$ & $0.7$ & $0.64$ & $5.2$ & $116$ \\
    {\tt RIT\_eBBH\_1639} & $111$ & $111$ & $272$ & $1$ & $0$ & $0.7$ & $0.75$ & $1.1$ & $113$  \\
    \hline
     \end{tabular}
     \caption{Parameters of the two highly eccentric NR injections used to assess the performance of {\tt TEOBResumS} in the high-eccentricity regime favored by GW190521. Extrinsic parameters and $M^\text{det}_{\text{total}}$ are determined by maximizing the likelihood against the GW190521 data. The corresponding $\ln \mathcal{L}_{\max}$ values are listed in the final column. Masses are given in $M_\odot$ and $D_L$ is reported in Gpc. }
    \label{tab:gw190521_NR_injection}
\end{table*}

\begin{figure*}
    \centering
    \includegraphics[scale=0.45]{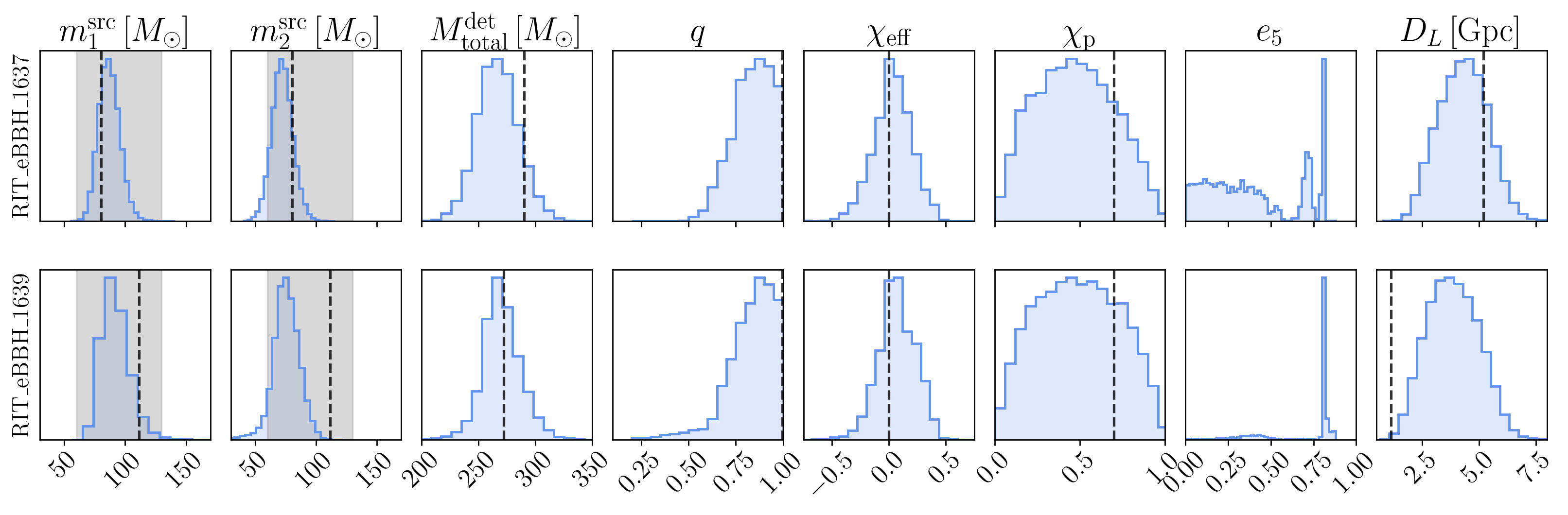}
    \caption{One-dimensional posterior distributions obtained by recovering two highly eccentric NR waveforms using {\tt TEOBResumS-E+P}. The dashed vertical lines indicate the injected values and the grey shaded region represents the PISN gap.  For clarity, the $e_5$ posterior is shown using a larger number of bins in order to resolve the sharp peak at high eccentricity. Note that frequency-dependent quantities for the NR injections are defined at the starting frequency of the NR waveform ($\sim2$\,Hz), whereas the posteriors are generated using a reference frequency of $5$\,Hz. Furthermore, due to differences in eccentricity conventions and reference frequencies, the injected eccentricity is not shown.}
    \label{fig:gw190521_NR_injections}
\end{figure*}

The highly eccentric interpretation of GW190521 presented in this work lies in a region of parameter space where eccentric waveform models have not been validated against NR waveforms. This motivates us to assess waveform-model performance in this regime and, consequently, the robustness of our inference. For this study, we focus on {\tt TEOBResumS}. While this model can generate waveforms up to eccentricities of $\sim0.9$, it has only been validated against NR waveforms up to moderate eccentricities \cite{Gamba_2022,PhysRevD.111.024069}, with a previous study finding good agreement between the model and NR up to initial eccentricities of $\sim0.7$ \cite{hwr5-scp4}.

To this end, we perform zero-noise injection--recovery tests using NR waveforms. Specifically, we inject highly eccentric NR waveforms and recover them using {\tt TEOBResumS-E+P}, assessing whether the injected parameters are recovered and whether the morphology of the injected waveform is reproduced. Our goal here is not to establish the overall accuracy of {\tt TEOBResumS}, but rather to evaluate its reliability for the specific highly eccentric configurations favored by GW190521.  
As such, we consider NR waveforms that were directly compared against GW190521 in Ref.~\cite{Gayathri2022}. Specifically, we focus on two waveforms from the RIT catalog \cite{Healy_2022} that were found to yield high likelihoods when compared against the GW190521 data: {\tt RIT\_eBBH\_1637} and {\tt RIT\_eBBH\_1639}.

For each waveform, we use the {\tt RIFT} framework to determine the extrinsic parameters and $M^\text{det}_\text{total}$ that maximize the likelihood against the GW190521 data. The parameters, together with the corresponding $\ln \mathcal{L}_{\max}$ values, are listed in Table~\ref{tab:gw190521_NR_injection}. The quoted eccentricities correspond to the initial values $e_0$, defined at instantaneous frequencies $\sim2$ Hz.  Furthermore, the simulations employ a different definition of eccentricity. In these simulations, the eccentricity is defined as $e = 2\epsilon - \epsilon^2$, where $\epsilon$ modifies the quasicircular tangential momentum $p_{\mathrm{t,qc}}$ according to
\begin{equation}
p_{\mathrm{t,ecc}} = p_{\mathrm{t,qc}}(1-\epsilon),
\end{equation}
where $p_{\mathrm{t,ecc}}$ is the eccentric tangential momentum used to generate the eccentric orbit and $0 < \epsilon < 1$. Because these NR waveforms only have a few cycles, we are unable to compute $e_5$ using {\tt gw\_eccentricity}. 

We inject each waveform into zero noise using the maximum-likelihood parameters listed in Table~\ref{tab:gw190521_NR_injection} and perform parameter estimation using the same settings as in the preceding analyses. The posterior distributions for select parameters are shown in Fig.~\ref{fig:gw190521_NR_injections}. We find that the injected masses, spin parameters, and $D_L$ are consistent with the recovered posterior distributions for both NR waveforms. However, for {\tt RIT\_eBBH\_1639}, the injected $D_L$ value, and consequently the inferred $m_2^\mathrm{src}$, lies near the edge of the recovered posterior. This performance is notable given that the injections probe a highly eccentric region of parameter space beyond that in which agreement between {\tt TEOBResumS-E+P} and NR waveforms has previously been demonstrated. Interpreting the recovered eccentricity is less straightforward, since the injected NR waveforms employ a different eccentricity definition and the quoted eccentricities correspond to initial values defined at different frequencies. We therefore refrain from making direct quantitative comparisons between the injected eccentricities and the recovered $e_5$ posteriors. Nevertheless, the results provide useful information regarding the ability of the model to capture highly eccentric waveform morphology and identify the highly eccentric nature of the signal.

\begin{figure*}
    \centering
    \includegraphics[scale=0.45]{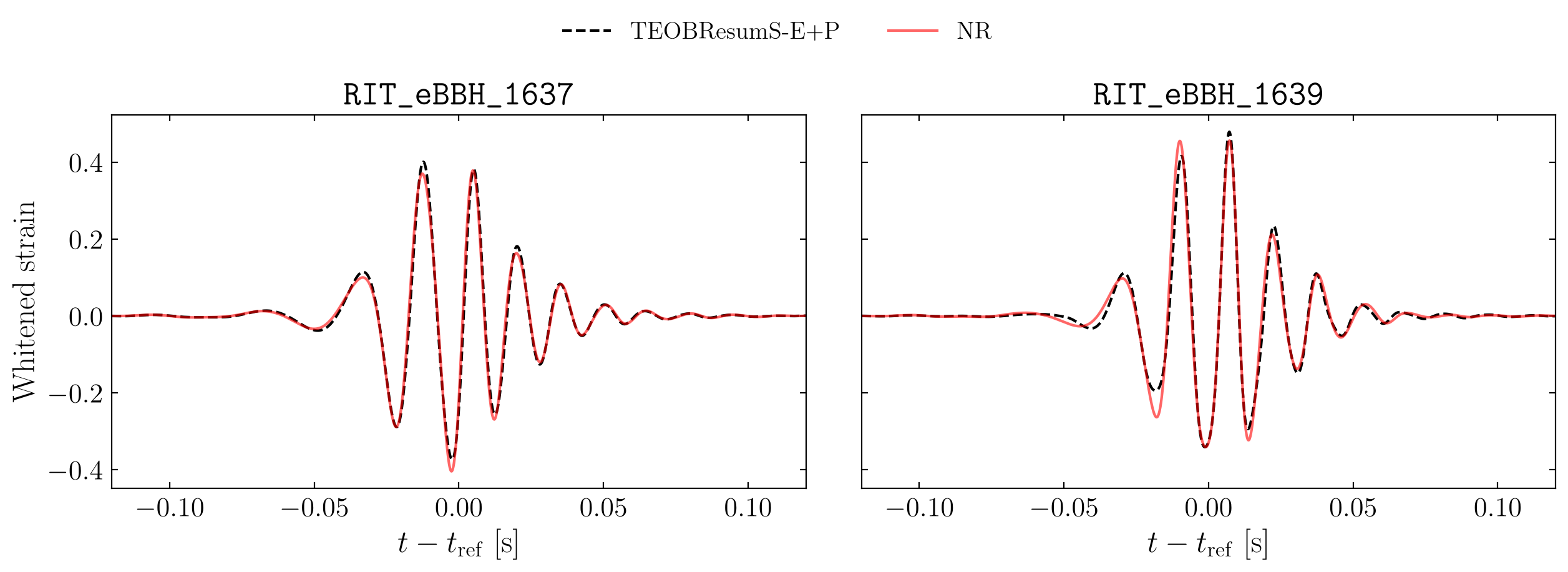}
    \caption{Comparison of the whitened LIGO Hanford strain for the NR injections with corresponding best-fit waveforms recovered using {\tt TEOBResumS-E+P}. The left and right panels show the RIT\_eBBH\_1637 and RIT\_eBBH\_1639 injections, respectively. Time is shown relative to $t_{\rm ref}$, defined as the time at which the magnitude of the complex-valued strain reaches its peak at the geocenter. In both cases, {\tt TEOBResumS-E+P} reproduces the overall morphology of the injected NR signal.}
    \label{fig:gw190521_NR_injections_morphology}
\end{figure*}

For both injections, we find {\tt TEOBResumS-E+P} is able to reproduce the overall morphology of the injected NR signal, as shown in Fig.~\ref{fig:gw190521_NR_injections_morphology}. The recovery of {\tt RIT\_eBBH\_1639} confidently favors nonzero eccentricity, although the $e_5$ posterior exhibits mild multimodal structure. In contrast, the recovery of {\tt RIT\_eBBH\_1637} exhibits substantially stronger multimodality, with significant posterior support extending toward lower eccentricities, including $e_5=0$. 
Thus, although the model can reproduce the injected waveform morphology in both cases, it does not always uniquely identify the signal as highly eccentric. This behavior may also contribute to the multimodal structure observed with {\tt TEOBResumS}, but not with {\tt IMRPhenomTEHM}, in the analysis of GW190521 presented in Sec.~\ref{ssec:gw190521_eccentricity}. We also note that the recovered {\tt TEOBResumS-E+P} posteriors favor eccentricities larger than the initial values quoted for the NR injections. Since $e_0$ is defined at $2\,$Hz and eccentricity is expected to decrease as the binary evolves, one would generally expect the corresponding eccentricity at $5\,$Hz to be smaller, although this expectation does not account for differences in the eccentricity definitions. This raises the possibility that the orbital eccentricity of GW190521 could be lower than the value inferred with {\tt TEOBResumS-E+P}.

The difference in eccentricity recovery between the two injections may arise from several factors. One possible contributing factor is the larger initial eccentricity of {\tt RIT\_eBBH\_1639}, which may produce waveform features that are more difficult to mimic using lower-eccentricity or quasicircular configurations, thereby reducing degeneracy in the recovery. Another possible explanation is the difference in the injected $M_\text{total}^\text{det}$ between the two systems. Because {\tt RIT\_eBBH\_1639} is injected at a lower $M_\text{total}^\text{det}$, the signal contains slightly more inspiral content within the detector band. Since eccentric effects are expected to be most prominent during the inspiral, the additional inspiral content may allow \texttt{TEOBResumS-E+P} to better reproduce the eccentric structure of the signal before it enters the merger-ringdown regime, where the waveform model follows quasicircular prescription. We also note that, because these NR waveforms contain only a small number of cycles, the injections were generated using minimal windowing, which could possibly be impacting the analysis. These explanations are  speculative and would require in-depth studies to determine which, if any, is responsible for the observed differences in the inference of the two injections.

Overall, these results indicate that \texttt{TEOBResumS-E+P} is capable of reproducing the features present in the highly eccentric NR waveforms and increase confidence in the robustness of our high-eccentricity inference, particularly given the recovery of {\tt RIT\_eBBH\_1639}. At the same time, the recovery of {\tt RIT\_eBBH\_1637} demonstrates that accurately reproducing the signal morphology does not necessarily lead to an unambiguous identification of the signal as highly eccentric, highlighting the challenges associated with eccentricity inference for short, highly eccentric signals. Finally, the large likelihoods obtained for the two highly eccentric NR waveforms when compared against the GW190521 data provide independent evidence that highly eccentric configurations can provide good agreement with the observed signal morphology.

\section{Conclusions}
In this work, we performed a detailed reanalysis of GW190521 to constrain its orbital eccentricity, assess the robustness of previous evidence for spin precession, and investigate the origin of the waveform-modeling systematics observed for this event. We analyzed the signal using multiple waveform families spanning eccentric aligned-spin, eccentric spin-precessing, and quasicircular spin-precessing configurations.

We found that analyses allowing eccentricities up to $e_5=0.9$ favor a highly eccentric interpretation of GW190521, with posterior support concentrated near $e_5\sim0.8$. Restricting the prior to $e_5\leq0.4$, however, prevents the analyses from accessing this high-likelihood region and yields uninformative eccentricity posteriors. Within the eccentric aligned-spin {\tt TEOBResumS} analysis, the full eccentricity range is favored over the restricted range by $\ln\mathrm{BF}\sim4.9$. These results demonstrate the importance of exploring sufficiently broad eccentricity ranges when analyzing GW190521, and how limited ranges could hide the highly eccentric interpretation. This further provides an explanation for why several previous analyses with more restrictive eccentricity prior ranges did not find evidence for eccentricity. 

We also find no indication of spin precession in the event when eccentricity is included in the analysis. While quasicircular spin-precessing models generally favor moderate-to-large $\chi_\mathrm{p}$, allowing both eccentricity and precession shifts the inference toward lower $\chi_\mathrm{p}$. This suggests that at least part of the apparent precession inferred under the quasicircular hypothesis may result from the models attempting to reproduce eccentric features in the signal. Furthermore, within the {\tt TEOBResumS} and {\tt IMRPhenomT} frameworks, the eccentric aligned-spin hypothesis is favored over the corresponding quasicircular spin-precessing hypothesis by $\ln\mathrm{BF}\sim2.6$ and $\sim4$, respectively. 

Our zero-noise injection studies provide further support for the conclusions drawn from our analysis of the real data. When a highly eccentric, non-precessing GW190521-like signal is analyzed with quasicircular spin-precessing models, we recover substantial support for spin precession. The injection studies also provide insight into the broader waveform-model systematics observed for GW190521: for the eccentric injection, the quasicircular waveform models recover different spin-precession, mass-ratio, and component-mass posteriors, reproducing the model-dependent behavior seen in the real data, whereas the quasicircular, highly spin-precessing injection yields broad agreement across waveform models. These results suggest that unmodeled eccentricity can contribute to both the apparent spin precession and the broader waveform-model systematics observed for GW190521.

Finally, using two highly eccentric NR waveforms previously compared with GW190521, we tested {\tt TEOBResumS-E+P} in the highly eccentric region favored by GW190521. For both injections, the model successfully reproduces the waveform morphology and broadly recovers the injected masses, spins, and distances. The eccentricity recovery, however, differs between the two cases: for one injection, the model correctly identifies the signal as highly eccentric, with strong posterior support at large $e_5$, whereas for the other, the recovered $e_5$ posterior is multimodal and includes substantial support near $e_5=0$.

Overall, our results favor a highly eccentric interpretation of GW190521 and demonstrate that accounting for eccentricity substantially changes the inferred properties of the source. In particular, eccentricity provides a consistent explanation for both the apparent spin precession inferred by quasicircular analyses and at least part of the waveform-model dependence of the inferred masses and spins. GW190521 therefore illustrates how restricting the physical content or parameter space of waveform models can lead to qualitatively different interpretations of short, high-mass gravitational-wave signals.

Caution is required when drawing astrophysical inferences about GW190521 from this study. Our Bayes factors, $\mathrm{BF} \sim 10$--$50$, in favor of an eccentric interpretation must be combined with astrophysical prior odds to determine the corresponding odds ratio. The expected eccentricity distribution depends on the binary formation channel \cite{Rodriguez_2018,PhysRevD.97.103014,Zevin_2019,Samsing:2020tda,Gondan:2020svr,Dorozsmai:2025jlu,Rozner:2026jtj}, and very high eccentricities are expected to be rare even among dynamically formed binaries. As the catalog of GW detections continues to grow in the coming years, our understanding of the formation channels of massive binary black holes will improve, enabling a more complete assessment of the astrophysical eccentricity distribution and, consequently, of the prior odds for an eccentric interpretation of GW190521.

\begin{acknowledgements}
We thank Maria de Lluc Planas for providing access to the {\tt IMRPhenomTEHM} waveform model and for a thorough review of the manuscript. We also thank Sylvia Biscoveanu, Juan Calderon Bustillo, Jacob Lange, Carlos Lousto, Lorenzo Pompili, and Harrison Siegel for their helpful comments. The posterior samples generated in our analysis, along with the configuration file required to reproduce our analysis, are available at \href{https://github.com/AasimZJan/GW190521_samples}{https://github.com/AasimZJan/GW190521$\_$samples}. A.J. and D.S. acknowledge support from NASA Grant No. 80NSSC24K0437 and NSF Grant No. PHY-2207780. 
A.Z. acknowledges support from NSF Grant PHY-2308833.
K.C. was supported by NSF Grant PHY-2409001.
R.O.S.  gratefully acknowledges support from NSF Award No. NSF PHY-1912632, No. PHY-2012057, No. PHY-2309172, No. AST-2206321, and the Simons Foundation. 
This material is based upon work supported by NSF's LIGO Laboratory, which is a major facility fully funded by the NSF. The authors are grateful for computational resources provided by the LIGO Laboratory and supported by NSF Grants PHY-0757058 and PHY-0823459.  
The work was done by members of the Weinberg Institute and has an identifier of UT-WI-26-2026.
\end{acknowledgements}

\bibliography{references}

\appendix

\section{Impact of excessive tapering on inference}
\label{app:excessive_taper}

To mitigate spectral leakage when transforming time-domain waveforms into the frequency domain, the beginning of each waveform is tapered. By default, {\tt RIFT} sets the taper duration to the larger of $1\%$ of the waveform duration and $1/f_{\min}$. For the short, highly eccentric waveforms relevant to GW190521, $1/f_{\min}$ is the larger of these two timescales and therefore determines the taper duration. This timescale approximates one waveform cycle when the frequency evolves slowly. However, for highly eccentric systems close to merger, the frequency can evolve substantially over $1/f_{\min}$, causing the standard prescription to taper a significant fraction of the waveform and remove appreciable signal power.

We find that this excessive tapering can significantly affect the inferred parameters. Using {\tt TEOBResumS-E} with our modified tapering prescription, where we strictly taper $1\%$ of the waveform, we infer $D_L=4.2^{+1.5}_{-1.6}$ Gpc, $m_1^\text{src}=96^{+23}_{-17}\,M_\odot$ and $m_2^\text{src}=73^{+17}_{-14}\,M_\odot$, as listed in Table~\ref{tab:gw190521_summary}, consistent with quasicircular spin-precessing models. In contrast, using the standard tapering prescription suppresses the waveform amplitude and shifts the inferred luminosity distance to a smaller value, $D_L=2.1^{+1.6}_{-1.1}$ Gpc. Because the detector-frame masses remain largely unchanged, the smaller inferred distance, and hence smaller cosmological redshift, shifts the source-frame component masses to larger values, $m_1^\text{src}=116^{+24}_{-23}\,M_\odot$ and $m_2^\text{src}=87^{+20}_{-19}\,M_\odot$. We also observe noticeable differences in the inferred anomaly parameter and $\chi_{1z}$. These results demonstrate that the tapering prescription can have a significant impact on parameter inference for very short, highly eccentric signals and must therefore be treated carefully in this regime.

\section{Impact of the $e_5$ prior upper bound on eccentricity inference}
\label{app:gw190521_e5_upper_limit}
\begin{table}
\centering
\renewcommand{\arraystretch}{1.4}
\addtolength{\tabcolsep}{4pt}
\begin{tabular}{| c | c | c |}
\hline
$e^{\max}_5$ & $e_5$ estimate & $\ln Z$ \\ [0.5ex]
\hline\hline

$0.5$ & $0.18^{+0.22}_{-0.16}$ & $77.3$ \\

$0.6$ & $0.18^{+0.23}_{-0.16}$ & $77.2$ \\

$0.7$ & $0.20^{+0.49}_{-0.18}$ & $77.2$ \\

$0.8$ & $0.72^{+0.03}_{-0.63}$ & $78.5$ \\

\hline
\end{tabular}
\caption{Inferred $e_5$ estimates as a function of the upper bound of the $e_5$ prior, along with $\ln Z$. An upper bound of $e_5^{\max} \gtrsim 0.8$ is required to recover strong support for nonzero eccentricity for GW190521.}
\label{tab:emax_injections}
\end{table}

In Sec.~\ref{ssec:gw190521_eccentricity}, we analyzed GW190521 using {\tt TEOBResumS-E} with two choices for the upper bound of the $e_5$ prior, $e_5^{\max} = 0.4$ and $0.9$. Here, we extend this study by varying the upper bound over $e_5^{\max} \in \{0.5, 0.6, 0.7, 0.8\}$ to determine the minimum value of $e_5^{\max}$ required to obtain strong evidence for nonzero eccentricity in this event. The inferred median values and $90\%$ credible intervals are summarized in Table~\ref{tab:emax_injections}.

For $e_5^{\max} = 0.5$ and $0.6$, the recovered posteriors retain strong support for $e_5 = 0$ and exhibit broad uncertainties. For $e_5^{\max} = 0.7$, the posterior resembles the ones obtained for lower $e_5^{\max}$ but shows a small mode near the upper limit, leading to a much larger upper limit as reported in Table~\ref{tab:emax_injections}. For $e_5^{\max} = 0.8$, the posterior shifts toward large eccentricity, yielding $e_5 \sim 0.72$ and exhibiting strong support for nonzero values. Furthermore, the posterior accumulates near the prior boundary, indicating that the inference is limited by the upper bound. However, a small amount of support is retained for lower eccentricities, including $e_5=0$. These results demonstrate that $e_5^{\max}  \gtrsim 0.8$ is required to recover strong support for high eccentricity for this event.

\section{Robustness of eccentricity inference to the choice of reference frequency}
\label{app:fmin_impact}
All eccentricity posteriors presented in the main text are measured at $5$ Hz. Here, we extend our analysis to lower reference frequencies by performing inference with $f_\text{min}=3.4$ Hz, which in {\tt TEOBResumS-E} also sets the frequency at which the eccentricity is defined. This allows us to assess the robustness of the nonzero eccentricity measurement with respect to the choice of reference frequency as well as $f_\text{min}$.

The choice of this frequency is guided by the relation between orbital frequency and eccentricity derived in Ref.~\cite{PhysRev.136.B1224}. Starting from the inferred eccentricity at $5$ Hz, we use this relation to estimate the expected eccentricity at lower frequencies, ensuring that the inferred values remain within the allowed range of $e \lesssim 0.9$ of the model. Specifically, we use
\begin{equation}
f \propto
\frac{(1 - e^2)^{3/2}}{e^{18/19} \left(1 + \frac{121}{304} e^2 \right)^{1305/2299}}~,
\end{equation}
which follows from the Peters evolution of eccentric binaries under gravitational radiation reaction.  At $3.4$ Hz, we continue to find strong support for nonzero eccentricity, with tightly constrained posteriors. In particular, we obtain $e_{3.4} = 0.85^{+0.01}_{-0.01}$.

\section{Reproducing the $q\sim0.1$ mode with {\tt IMRPhenomXPHM}}
Here, we vary the analysis settings to test whether the $q \sim 0.1$ mode reported in Ref.~\cite{Nitz_2021} using {\tt IMRPhenomXPHM} arises from differences in analysis choices. To this end, we repeat the analysis using a configuration that closely follows theirs: we compute the likelihood over the frequency range $20\text{--}500$\,Hz with $f_{\text{min}} = 20$\,Hz. In the likelihood computations, we utilize the PSD that we estimate using the Welch method, performing a median--mean average over $71$ segments of $8$ seconds each, centered on the merger time, using the PyCBC package \cite{Biwer:2018osg}. We also increase the sampling rate to $1024$\,Hz and data segment length to $8$\,s. 

We find non-trivial differences in the resulting posteriors. Among intrinsic parameters, the largest differences are observed in $m_2^{\text{det}}$, and $\chi_{1z}$. All extrinsic parameters are also affected. The Bayesian evidence decreases from $\ln Z = 80.3$ to $76.1$. Despite these changes, we do not recover a third mode near $q \sim 0.1$. Interestingly, when restricting the mode content of the recovery waveforms to $l_{\text{max}} = 2$, the feature near $q \sim 0.2$ becomes somewhat more pronounced relative to higher mass ratios ($q \sim 0.4\text{--}1.0$). However, this enhancement is modest, and no distinct or statistically significant mode near $q \sim 0.1$ is recovered.

\end{document}